\documentclass[10pt,conference]{IEEEtran}
\IEEEoverridecommandlockouts

\usepackage{amsmath,amssymb,amsfonts, amsthm}

\usepackage{siunitx}
\usepackage{cite}
\usepackage{tikz}
\usetikzlibrary{automata, positioning, arrows,calc}
\usetikzlibrary{quantikz}
\usepackage{algorithmic}
\usepackage{graphicx}
\usepackage{textcomp}
\usepackage{xcolor}
\usepackage{caption} 
\usepackage{subcaption}
\usepackage{hyperref}
\usepackage{mathtools}
\usepackage{multirow}
\usepackage{graphicx}
\usepackage{multicol}
\usepackage{array}
\usepackage{mathrsfs}
\newtheorem*{remark}{Remark}
\usepackage{adjustbox}
\usepackage{latexsym}
\usepackage{tcolorbox}
\tcbuselibrary{breakable}
\usepackage{afterpage}
\usepackage{ragged2e} 
\usetikzlibrary{trees}
\usepackage{cuted}
\usepackage{bigstrut}
\usepackage{makecell}
\usepackage{booktabs}
\usepackage{tabularx}
\usepackage[utf8]{inputenc}

\renewcommand{\arraystretch}{1.3}

\begin{document}

\title{Quantum Transduction:\\
Enabling Quantum Networking
\thanks{}
}

\makeatletter
\newcommand{\linebreakand}{%
  \end{@IEEEauthorhalign}
  \hfill\mbox{}\par
  \mbox{}\hfill\begin{@IEEEauthorhalign}
}
\makeatother

\author{
    \IEEEauthorblockN{Marcello Caleffi$^{\dagger}$,~\IEEEmembership{Senior~Member,~IEEE}, Laura d'Avossa$^{\dagger}$,~\IEEEmembership{Student~Member,~IEEE}, \\
    Xu Han$^{*}$, Angela Sara Cacciapuoti$^{\dagger}$,~\IEEEmembership{Senior~Member,~IEEE},}
    \thanks{$^\dagger$The authors are with the \href{www.quantuminternet.it}{www.QuantumInternet.it} research group, University of Naples Federico II, Naples, 80125 Italy.}
    \thanks{$^*$The author is with the Center for Nanoscale Materials, Argonne National Laboratory, Lemont, Illinois 60439, USA.}
    \thanks{Corresponding author: Marcello Caleffi. Marcello Caleffi and Laura d'Avossa contributed equally to this work. This work has been funded by the European Union under Horizon Europe ERC-CoG grant QNattyNet, n.101169850. Views and opinions expressed are however those of the author(s) only and do not necessarily reflect those of the European Union or the European Research Council Executive Agency. Neither the European Union nor the granting authority can be held responsible for them. The work has been also partially supported by PNRR MUR RESTART-PE00000001. Work performed at the Center for Nanoscale Materials, a U.S. Department of Energy Office of Science User Facility, was supported by the U.S. DOE, Office of Basic Energy Sciences, under Contract No. DE-AC02-06CH11357. Xu Han acknowledges support from the Argonne National Laboratory Directed Research and Development (LDRD) program, and support from the U.S. DOE, Office of Science, Advanced Scientific Computing Research (ASCR) program under Contract No. DE-AC02-06CH11357 as part of the InterQnet quantum networking project.
    }
}

\maketitle

\begin{abstract}
The complementary features of different qubit platforms for computing and communicating impose an intrinsic hardware heterogeneity in any quantum network, where nodes, while processing and storing quantum information, must also communicate through quantum links. Indeed, one of the most promising hardware platforms at quantum node scale for scalable and fast quantum computing is the superconducting technology, which operates at microwave frequencies. Whereas, for communicating at distances of practical interest beyond few meters, quantum links should operate at optical frequencies. Therefore, to allow the interaction between superconducting and photonic technologies, a quantum interface, known as \textit{quantum transducer}, able to convert one type of qubit to another is required. 
This paper aims to provide a tutorial treatise on the fundamental research challenges of quantum transduction. The tutorial is structured around a communications engineering framework, thereby shedding light on its fundamental role in quantum network design and deployment—a perspective often overlooked in existing literature. This framework allows us to categorize different transduction modalities and to reveal an unorthodox one where the transducer itself can act as an entanglement source.
From this standpoint, it is possible to conceive different source-destination link archetypes, where transduction plays a crucial role in the communication performances. The analysis also translates the quantum transduction process into a proper functional block within a new communication system model for a quantum network.
\end{abstract}


\section{Introduction}
\label{sec:01}

The scientific and industrial communities recognize the imperative to use different technologies to achieve the ultimate vision of the Quantum Internet, as there is no single hardware platform that can address all the challenges connected to \textit{store}, \textit{process} and \textit{communicate} quantum states \cite{CacCalTaf-20, DavCacCal-24, DavCalWan-23, DavCacCal-2024, DavZhaChu-24}. Indeed, there exist several hardware platforms for realizing a quantum bit (qubit), and each of them exhibits different advantages and limitations.
Although an overview of the state-of-the-art of qubit hardware platforms and their main features is summarized in Table~\ref{tab:01} \cite{McKinsey2025QuantumMonitor}, in this work we focus on superconducting and photonic technologies. The rationale is that, as at the time of writing, these two platforms are the ones most developed and more widely accepted by the scientific community for the implementation of computational and flying qubits \cite{IllCalMan-22}.

On one hand, \textit{superconducting technology} stands out as one of the most promising platforms for universal quantum computing. Indeed, superconducting quantum circuits are characterized by high-scalability and ``fast'' gates \cite{KjaSchBra-20, DavWalMar-04, AruaryBab-19}. For these reasons, they represent the hardware underlying computational qubits chosen by most of the major players, such as IBM, Google, Rigetti, Fujitsu and Alice$\&$Bob. With superconducting qubits, logical operations can be precisely controlled through carefully shaped (typically, microwave) pulses that alter the electrodynamics and energy levels of the qubit. 
Furthermore, since superconducting qubits are realized using techniques similar to those of classical semiconductor chips, they appear an ideal technology for fabricating quantum processors with easily scalable numbers of qubits. For instance, the latest IBM processor has already reached over 1000 physical qubits and the company plans to achieve 2000 logical qubits by 2033 \cite{McKinsey2025QuantumMonitor, IBM2}.
Yet, superconducting qubits require challenging cryogenic temperatures (in the order of millikelvin) through dilution refrigerators. It has been proved that different superconducting processors can be connected through cryogenic waveguides \cite{XiaZhaLia-17, KurPechRoy-19, MagStoKur-20}. However, such cryogenic cables introduce severe communication constraints, limiting both communication ranges and achievable network topologies. Furthermore, they have very expensive manufacturing costs and non-trivial installation requirements. Thus, they are not suitable for the realization of even small-to-moderate-size quantum networks \cite{CacPelIll-25}.

\begin{table*}[h]
\centering
\resizebox{\textwidth}{!}{%

\begin{tabular}{|>{\centering\arraybackslash}m{2.5cm}
                |>{\centering\arraybackslash}m{2.5cm}
                |>{\centering\arraybackslash}m{2.5cm}
                |>{\centering\arraybackslash}m{2.5cm}
                |>{\centering\arraybackslash}m{2.5cm}
                |>{\centering\arraybackslash}m{2.5cm}|}
\hline\hline
\textbf{Technology} & \textbf{Optical photons} & \textbf{Superconducting circuits} & \textbf{Spin qubits} & \textbf{Neutral atoms} & \textbf{Trapped ions} \\
\hline\hline
\textbf{Qubit encoding} 
    & Polarization, time-bin or path of single photons
    & Energy states of Cooper pairs across Josephson junctions
    & Electron spins in semiconductor quantum dots or donor atoms
    & Internal electronic states of atoms trapped by optical tweezers or lattices; entangling via Rydberg blockade
    & Internal electronic states of ions confined by electromagnetic field; entangling via shared motional modes\\
\hline
\textbf{Physical qubit count} & 216 - Xanadu
 & 1.121 – IBM \newline 105 - Google \newline 84 - Righetti \newline 64 - Fujitsu \newline 16 - Alice $\&$ Bob & 12 - Intel & 1.600 – Infleqtion \newline
1.180 - Atom Computing
\newline
1.110 - Pasqal
\newline
256 - QuEra
 & 56 - Quantinuum
 \newline
36 – IonQ
\newline
20 - Alpine Quantum Technologies
 \\
\hline
\textbf{Control $\&$ Readout} & Integrated photonic circuits: beam splitters, phase shifters, polarization control, single-photon detectors
    & Microwave pulses for single- and two- qubit gates; dispersive readout with resonators
    & Electrostatic gate control + microwave/ electric pulses for spin rotations; spin-dependent tunnelling readout &  Laser beams for trapping an addressing; fluorescence imaging for readout
 & Laser beams (single- and two-qubit gates); fluorescence readout; microwave-assisted gates in some approaches
 \\
\hline
\textbf{Cryogenic control} & Not required & Required & Required & Some solutions & Some solutions \\
\hline
\textbf{Main challenges} & Develop deterministic, high-fidelity and fast two-qubit gates & Integration of control electronics inside cryostat to meet performance demands & Adapting single-flux-quantum control chips to spin qubits  & Develop higher-powered lasers for individual qubit control while reducing photon loss
 &  Optimize space in vacuum chamber for having high number of individually controlled qubits \\
\hline
\textbf{Pros} & Room-temperature operation; long-distance transmission via fiber; excellent for quantum networking & Fast gates; strong industrial support; mature fabrication techniques & CMOS compatibility; potential for very high density; long coherence in isotopically pure materials & Naturally identical qubits; large arrays possible; long coherence times & Highest gate fidelities demonstrated; long coherence times; identical qubits\\
\hline
\textbf{Cons} & Lack of deterministic entangling gates; photon loss limits scalability & Requires dilution refrigerators; coherence times still limited; wiring bottlenecks & Fabrication variability; two-qubit gate fidelity still challenging; cryogenics required & Laser complexity; gate errors from Rydberg decay; optical alignment scalability & Slow gate speeds; scaling beyond tens-hundreds of ions is difficult; laser overhead
\\
\hline\hline
\end{tabular}
} 
\caption{Overview of qubit modalities and their main features.}
\label{tab:01}
\end{table*}

On the other hand, \textit{photonic technology} is worldwide recognized as the ``technology'' for communication purposes. Indeed, weak interaction with the environment (thus, reduced decoherence), low-loss transmissions, easy control with standard optical components, and lowest transmission delays make optical photons the ideal candidates to interconnect remote quantum processors \cite{RenXuYon-17,CacCalTaf-20}. Therefore, there exists a broad consensus within the research community about optical photons being the most promising platform for implementing \textit{flying qubits}. Indeed, quantum communication using optical photons has been demonstrated across long distances, over $400$Km via optical fibers \cite{YanLiuFen-2025} and $1000$Km in free-space \cite{JiPinLin-17, JuaYuaYuH-17}, respectively. Unfortunately, optical photons are not the ideal candidate when it comes to quantum computing. Indeed, although single-qubit quantum gates can be performed through optical components such as wave-plates \cite{BouFenDun-24, CalAmoFer-22}, optical photons do not naturally interact with each other, making yet very challenging to develop deterministic, high-fidelity and fast two-qubit gates \cite{KniLafMil-01}.

\begin{figure}[t!]
    \centering
    \includegraphics[width=1\columnwidth]{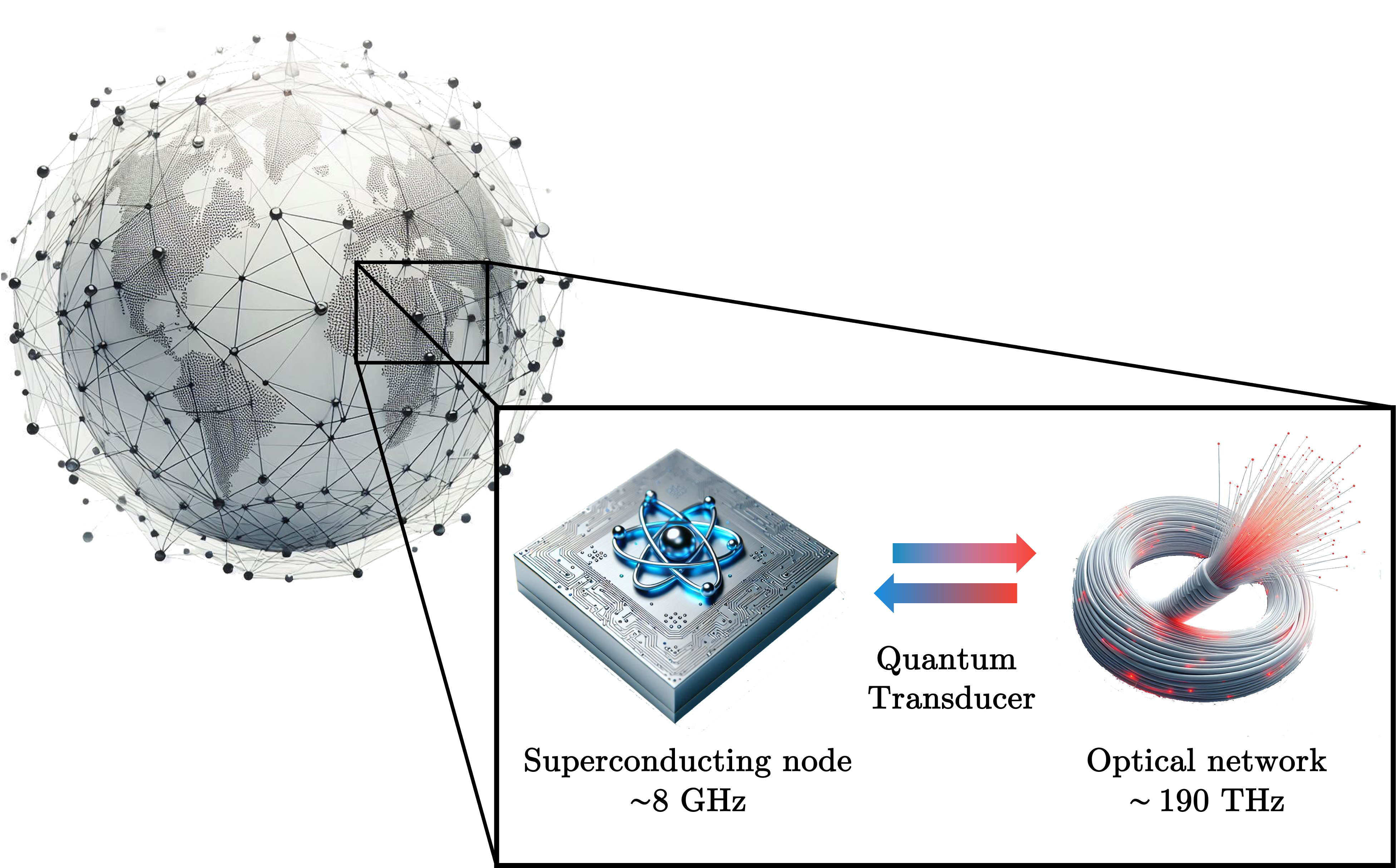}
    \caption{Schematic representation of a \textit{quantum transducer} as an interface between superconducting quantum nodes and optical quantum links. As highlighted within the figure, the frequency gap between microwave and optical frequencies spans five orders of magnitude, making the transduction between the two hardware platforms one of the most challenging nowadays \cite{LauSinBar-20}.}
   \label{fig:01}
   \hrulefill
\end{figure}

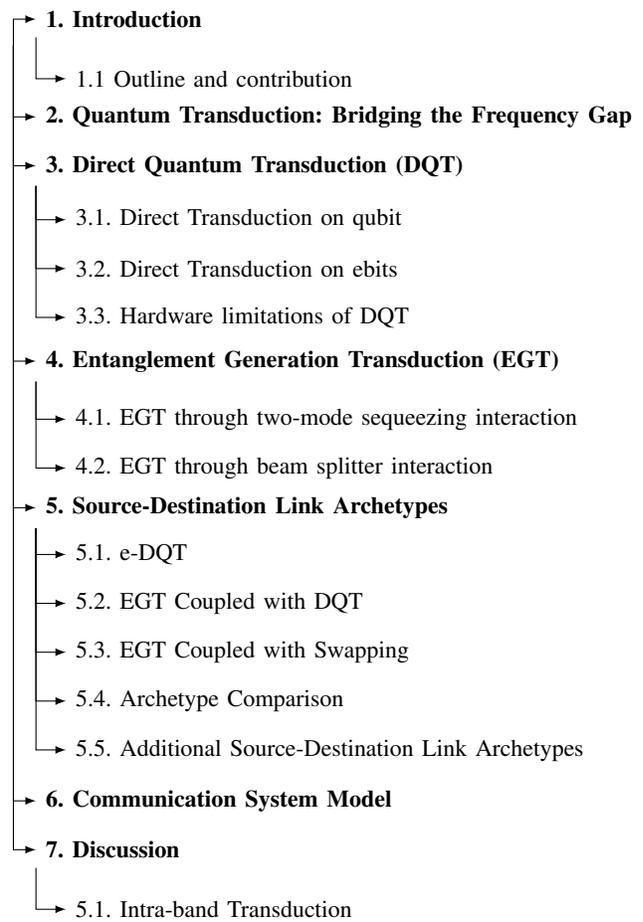
\begin{figure}[t]
    \centering
    \tikzstyle{every node}=[thick,anchor=west]
    \begin{tikzpicture}[level distance=2cm, grow via three points={one child at (0.3,-0.55) and two children at (0.3,-0.5) and (0.3,-1.15)}, edge from parent path={([xshift=0.0mm] \tikzparentnode.south west) |- (\tikzchildnode.west)}, growth parent anchor=south west, edge from parent/.style = {draw, -latex}]
        \node {}
        child {
            node {\small \textbf{1. Introduction}}
            child[xshift=0.1cm] { node {\small 1.1 Outline and contribution}}
        }		
        child [missing] {}
        child { 
        node[align=left, text width=8cm] {\small \textbf{{2. Quantum Transduction: beyond classical conversion}}}
        }
        child {
            node {\small \textbf{3. Direct Quantum Transduction (DQT)}}
            child[xshift=0.1cm] { node {\small 3.1. Direct Transduction of Information Carriers}}
            child[xshift=0.1cm] { node {\small 3.2. Direct Transduction of Entanglement Carriers}}
            child[xshift=0.1cm] { node {\small  3.3. Conversion efficiency}}
            child[xshift=0.1cm] { node {\small  3.4. Quantum Channel Capacity}}
        }
        child [missing] {}
        child [missing] {}
        child [missing] {}
        child [missing] {}
        child {
            node {\small \textbf{4. Entanglement Generation Transduction (EGT)}}
            child[xshift=0.1cm] { node {\small 4.1. EGT through two-mode squeezing interaction}}
            child[xshift=0.1cm] { node {\small 4.2. EGT through beam-splitter interaction}}
            child[xshift=0.1cm] { node {\small 4.3. Quantum Channel Capacity}}
        }
        child [missing] {}				
        child [missing] {}
        child [missing] {}
        child {
            node {\small \textbf{5. Source-Destination Link Archetypes}}
            child [xshift=0.1cm] { node {\small 5.1. e-DQT}}
            child [xshift=0.1cm] { node  {\small 5.2. EGT Coupled with DQT}}
            child[xshift=0.1cm] { node {\small 5.3. EGT Coupled with Swapping}}
            child[xshift=0.1cm] { node {\small 5.4. Archetypes Comparison}}
            child[xshift=0.1cm] { node {\small 5.5. Additional Source-Destination Link Archetypes}}
        }
        child [missing] {}				
        child [missing] {}				
        child [missing] {}
        child [missing] {}
        child [missing] {}
        child {
            node {\small \textbf{6. Communication System Model}}
        }
        child {
            node {\small \textbf{7. Discussion}}
            child[xshift=0.1cm] { node {\small 7.1 Intra-band Transduction}}
            child[xshift=0.1cm] { node {\small 7.2 Challenges and Future Directions}}
        }
        child [missing] {}
        child [missing] {}
        ;
    \end{tikzpicture}
    \caption{Paper Structure.}
    \label{fig:02}
    \hrulefill
\end{figure}

The complementary features of the two aforementioned quantum-hardware-platforms make them ideal candidates to fulfil the \textit{DiVincenzo criteria} for quantum computing and communication \cite{DiV-00}, thus fully unleashing the ultimate vision of the quantum revolution represented by the Quantum Internet. But, at the same time, these complementary features call for an inherent heterogeneity within a quantum network, where quantum states are processed by superconducting nodes and transmitted via flying qubits through optical quantum channels, as schematically depicted in Fig.~\ref{fig:01}.

In order to embrace and sustain such a heterogeneity within a quantum network, \textit{quantum transduction} is needed \cite{DavCacCal-2024, DavCacCal-24}. By oversimplifying, quantum transduction is the process of converting one type of qubit into another, thus making possible the interaction between superconducting and photonic hardware technologies, which unfortunately operate at extremely different energy scales \cite{LauSinBar-20}.

Indeed, flying qubits working at optical frequencies (typically about hundreds of terahertz) cannot directly interact with superconducting qubits, which conversely work at microwave frequencies (typically few GHz). Therefore, a quantum transducer is needed to convert the state of a superconducting qubit into the state of a flying qubit and vice-versa, by bridging the five-order frequency gap between microwave and optical frequencies and, at the same time, by preserving the quantum state from one form to another. To this aim, it is worthwhile to underline that unconventional quantum mechanics features -- such as the measurement postulate and the no-cloning theorem -- makes quantum transduction fundamentally different from classical transduction. In simple words, quantum transduction must obey the no-cloning theorem, meaning it cannot simply copy and amplify signals the way classical transducers do. And since any measurement would irreversibly collapse the quantum state, quantum transduction must perform a coherent unitary process able to map one quantum carrier into another without “looking” at it.

Accordingly, \textit{Quantum Transducers} (QTs) -- i.e., the network elements performing quantum transduction -- play a crucial role in any quantum network, by constituting a matter-flying interface capable of integrating different qubits platforms \cite{CalCacBia-18, LauSinBar-20, LamRueSed-20, RueSedCol-16, HeaRueSah-20}. Furthermore, the heterogeneity that must be handled by quantum transduction is not limited to different qubit platforms. As a matter of fact, there exist different physical channels through which flying qubits can be transmitted, ranging from free-space optical channels to optical fibers. As a result, the transduction should be designed by taking into account the peculiarities of the physical channel the flying qubits propagate through \cite{CacCalTaf-20}.

\subsection{Outline and contribution}
\label{sec:01.1}

In the last decade, the field of quantum transduction has advanced significantly from a hardware standpoint \cite{LauSinBar-20, HanFuZou-21, Lambert2019}. The physics and hardware-engineering communities have been active in investigating schemes and technologies enabling such an interface \cite{LauSinBar-20, XinYifChe-24}, with multiple solutions. Yet, numerous challenges still encompass the hardware realization of a quantum transducer, while the network architectures required for its integration are even more demanding, rendering both domains highly active areas of research.

In this context, the aim of this paper is to model quantum transduction from a complementary yet fundamental perspective, namely, from the \textit{communication engineering perspective}. In a nutshell, through the manuscript:
\begin{itemize}
    \item We introduce and analyze the two distinctive functional-operation-modes of a quantum transducer, namely, \textit{direct conversion} vs \textit{entanglement generation}.
    \item We investigate and discuss the role of quantum transduction in different network archetypes for quantum information transmission via quantum teleportation.
    \item We translate quantum transduction as a functional block of the quantum communication system model. Specifically, by resorting to a classical communication terminology, the transducer functionality is reminiscent of the modulator (transmission side) and de-modulator (receiver side).
\end{itemize}
Through these key contributions, the aim of this paper is to drive the reader, with a tutorial approach, to grasp  the \textit{fundamental research challenges underlying quantum transduction within the communication engineering domain}. To the best of authors’ knowledge, a tutorial of this type is the first of its own.

\begin{table*}[h]
    \centering
    \resizebox{\textwidth}{!}{%
        \begin{tabular}{|>{\arraybackslash}m{2.5cm}
                        |>{\arraybackslash}m{2.5cm}
                        |>{\arraybackslash}m{2.5cm}
                        |>{\arraybackslash}m{2.5cm}
                        |>{\arraybackslash}m{2.5cm}|}
            \hline\hline 
            \centering{\textbf{Technology}} 
            & \multicolumn{2}{|c|}{\centering{\makecell{\includegraphics[width=0.1\textwidth]{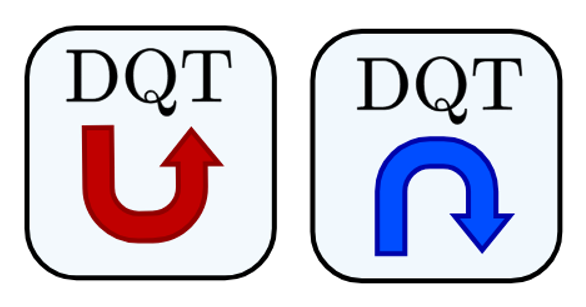} \\ \textbf{Direct Quantum Transduction (DQT)}}}} 
            & \multicolumn{2}{|c|}{\makecell{\includegraphics[width=0.05\textwidth]{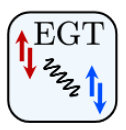} \\ \textbf{Entanglement Generation Transduction (EGT)}}} \\
            \hline
             & Qubits:  Sec.~\ref{sec:03.1} & Ebits:  Sec.~\ref{sec:03.2} & \multicolumn{2}{|c|}{Sec.~\ref{sec:04}} \\
            \hline\hline
            \textbf{Relevant references} & \multicolumn{2}{|c|}{\cite{Tsa-10, Tsa-11, HanFuZou-21}}  & \multicolumn{2}{|c|}{\cite{ZhoHanXu-20, CheZhoWan-22, Sahu2023}} \\
            \hline
            \textbf{Functionality} & Converts \textit{informational qubits} from one frequency domain to another
             & Converts \textit{ebits} from one frequency domain to another & \multicolumn{2}{|c|}{Generates hybrid entanglement between different frequency domains} \\
            \hline
            \textbf{Lossy Scenarios} & Whenever the quantum information is lost, it cannot be recovered & \multicolumn{3}{|c|}{Entanglement resource can be regenerated} \\
            \hline
            \textbf{Physical interaction exploited} & \multicolumn{2}{|c|}{Beam-splitter interaction} & Beam-splitter interaction with microwave filed initialization
             & Two-mode squeezing interaction \\
            \hline
            \textbf{Quantum Information transmission} & Direct transmission of quantum information
             & \multicolumn{3}{|c|}{Quantum information is transmitted through Quantum Teleportation}
             \\
            \hline
            \textbf{Quantum Capacity Metric} & One-way Quantum Capacity
             & \multicolumn{3}{|c|}{Two-way Quantum Capacity
            } \\
            \hline
            \textbf{Condition for no-null Capacity} & $\eta_\uparrow\eta_\downarrow>\frac{1}{2}$ & \multicolumn{3}{|c|}{$\eta_\uparrow\eta_\downarrow>0$} \\
            \hline
            \textbf{Role in quantum communication system} & \multicolumn{2}{|c|}{Direct modulation/demodulation} & \multicolumn{2}{|c|}{Un-direct modulation} \\
            \hline\hline
        \end{tabular}
    }
    \caption{Schematic summary of the two different operational modes of a quantum transducer: \textit{Direct Quantum Transduction} (DQT) vs \textit{Entanglement Generation Transduction} (EGT). DQT is further specialized depending on the type of transduced carrier: quantum information carriers (qubits) vs entanglement carriers (ebits).}
    \label{tab:03}
\end{table*}

The paper is structured as depicted in Fig.~\ref{fig:02}. Specifically, in Section~\ref{sec:02}, we introduce quantum transduction, highlighting its fundamentally different functionality and challenges compared to classical transduction. Moreover, we describe how quantum transduction can operate either on quantum-information carrier or on entanglement carrier for entanglement generation and distribution\footnote{We suggest an unfamiliar reader to refer to the Boxes named \textsc{Entanglement} and \textsc{Quantum Teleportation} to grasp the importance of entanglement as a communication resource, before delving into the field of quantum transduction acting on entanglement carrier. For a condensed and compact overview of quantum computing and quantum communication we refer the reader to \cite{KouCacCal-22, ZhoAnqNi-25, PanLonYin-24, SinDevSij-21, CacCalVan-20, IllCalMan-22}, while a more rigorous introduction can be found in \cite{nielsen00, RiefPol-11}.}.
Finally, we show that the same hardware device enabling transduction of quantum-information carriers (or entanglement carriers) -- namely, for the so-called \textit{direct} transduction -- can be remarkably used as an entanglement source. This allows us to distinguish\footnote{It is worthwhile to note that the jargon \textit{direct quantum transduction} is mainly limited by the literature to the process of transducing quantum information carrier \cite{HanFuZou-21, ZhoHanJia-22}, whereas in the following we extend its use to the transduction of entanglement carrier as well, being entanglement the fundamental resource of quantum networks \cite{KozWehVan-23}. Furthermore, in the same works, EGT coupled  with \textit{quantum teleportation} is also referred to as Entanglement-Based Quantum Transduction (EQT).} between \textit{Direct Quantum Transduction} (DQT) and \textit{Entanglement Generation Transduction} (EGT).

\begin{figure*}[h]
    \centerline{\includegraphics[width=\textwidth]{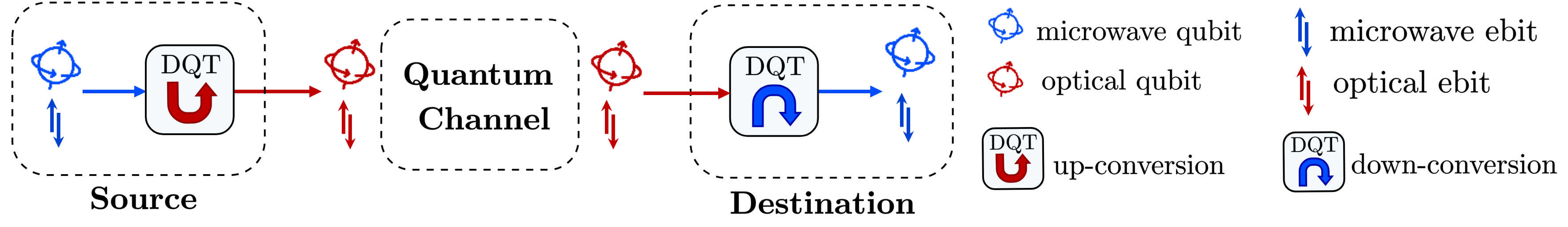}}
    \caption{Schematic representation of the interconnection of two superconducting quantum nodes via \textit{Direct Quantum Transduction} (DQT), converting either informational qubits or ebits. Microwave (optical) qubit and ebit are depicted in blue (red).}
    \hrulefill
    \label{fig:03}
\end{figure*}

In Section~\ref{sec:03}, we unpack the specifics of DQT, by discussing individually the two cases: the conversion of a quantum-information carrier (i.e., a qubit) in Section~\ref{sec:03.1} and the conversion of an entanglement carrier (i.e., an ebit) in Section~\ref{sec:03.2}, respectively. This section is enriched with an in-depth analysis of the performance of DQT in terms of the key performance indicator from a communication engineering perspective -- namely, the quantum channel capacity -- by taking into account the current state-of-the-art of transducer hardware. Through this analysis, we are able to highlight the reasons for which DQT acting on ebits is preferable with respect to DQT acting on informational qubits.

In Section~\ref{sec:04} we delve into the details of EGT, by introducing and discussing the ability of the transducer hardware to enable entanglement generation between the microwave and the optical domains. Specifically, we present two different physical interactions that can be exploited for entanglement generation: the \textit{two-mode squeezing} interaction in Section~\ref{sec:04.1} and the \textit{beam-splitter} interaction in Section~\ref{sec:04.2}, respectively.

Then, stemming on the material presented so far, in Section~\ref{sec:05} we provide some guidelines for elucidating and analysing how transduction is exploited for quantum information transmission. Specifically, we deepen different source-destination link archetypes, by exploiting different QT techniques. Our objective is to configure QT into network architecture considerations for a more comprehensive overview.

In Section~\ref{sec:06}, we introduce the transducer within the communication system model, designing it as a modulator/demodulator block. Counter-intuitively, while in the classical world there exists only one scheme for implementing modulation/demodulation -- namely, \textit{direct} modulation -- in a quantum network direct modulation is one possibility, indeed not even the most promising one due to the state-of-the-art limitations of transducer hardware. Therefore, we introduce the concept of \textit{un-direct modulation/demodulation.}

Finally, in Section ~\ref{sec:07}, we provide an outlook of the non-idealities affecting the transduction process and we discuss the open challenges, including the transduction between photons in the same frequency domain, namely, the so-called intra-band transduction.

\begin{figure*}[h]
    \centerline{\includegraphics[width=\textwidth]{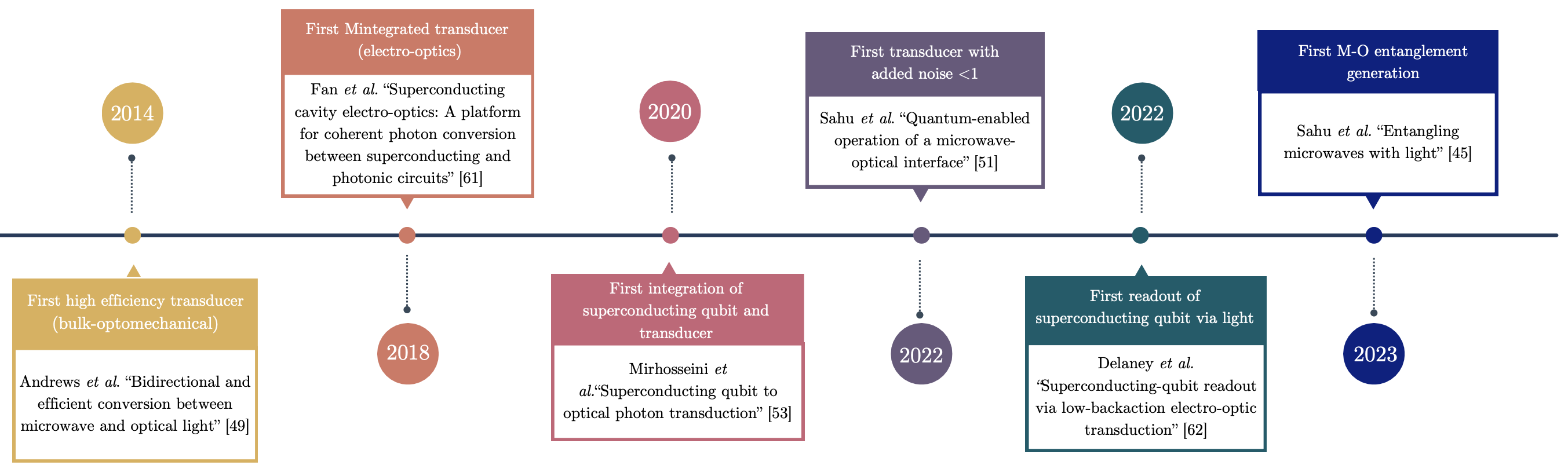}}
    \caption{Some key literature highlighting the major advancements and milestones in microwave-optical quantum transduction.}
    \hrulefill
    \label{fig:milestones}
\end{figure*}

\section{Quantum Transduction: \\
beyond classical conversion}
\label{sec:02}

In the classical domain, a transducer is required to convert the output of the information source into a signal (by oversimplifying) suitable for the transmitter functional block of a classical communication system \cite{Shan-48}. At the destination, a similar transducer is required to convert the propagated signal that is received into a form that is suitable for the user \cite{ProMas-01}.

This functionality relies on the ability of reading the informational content stored within the information carrier and rewriting it into another carrier of a different ``domain'' through a copy. Crucially, the physical laws underlying quantum transduction change in the quantum domain: due to the no-cloning theorem, the writing/reading via a copy becomes impossible. This highlights the most significant difference between classical and quantum transduction: although the two concepts share a similar aim, they are fundamentally different from an operational perspective. Indeed, the limitations of classical transduction are explicitly defined by the read-and-write processes. And current hardware technologies are now mature enough to widely overcome the technical challenges imposed by these processes. In other words, the implementation of classical transduction is now so consolidated that it no longer represents a major challenge from a communication perspective. On the other hand, the limitations of quantum transduction remain largely challenging. The key requirement is to preserve the coherence of the quantum state during the transfer from one domain to another without collapsing the original quantum state via measurement nor violating the no-cloning theorem.

Here we introduce an unique feature that characterizes quantum transduction: the ``nature'' of the carrier undergoing transduction. Similarly to classical transduction, a quantum transducer operates on a quantum information carrier, such as a qubit. But, differently from classical transduction, a quantum transducer can equivalently operate on an \textit{entanglement carrier}, namely, a carrier of entanglement encoding quantum correlation rather than quantum information such as an \textit{entanglement bit} (ebit)\footnote{In the following, \textit{EPR pair} and \textit{EPR} are used equivalently to denote a pair of maximally entangled qubits, and each qubit of the pair (with a slight abuse of notation, though, given that ebit is a unit of measure for entanglement) is referred to as \textit{ebit}.}, as shown in Table~\ref{tab:03}. In other words, a quantum transducer can operate either as frequency converter for an information carrier or as entanglement distributor for an entanglement resource \cite{CacCalVan-20}, i.e., an interface to transfer entanglement resources from one hardware platform to another.

But there exists a second key aspect characterizing quantum transduction and summarized again in Table~\ref{tab:03}: whether the hardware is used to transduce a carrier (as discussed above) or to directly generate hybrid entanglement, i.e. entanglement between the microwave and the optical domain. The two different operation-modes are thus referred to as Direct Quantum Transduction (DQT) and Entanglement Generation Transduction (EGT), respectively.

It is worthwhile to note that the difference between the two transduction-modes of DQT -- quantum information carrier vs entanglement carrier -- as well as the overall difference between the two operational-modes -- DQT vs EGT -- are very far from being only conceptual. Rather, they have profound impact from a communication engineering perspective, as summarized in Table~\ref{tab:03} and deeply discussed in Sections~\ref{sec:03} and \ref{sec:04}, respectively.

\section{Direct Quantum Transduction (DQT)}
\label{sec:03}

As introduced, DQT coherently converts superconducting qubits/ebits to flying qubits/ebits and vice versa, by allowing the  transmission of quantum information and entanglement resources among distant quantum nodes \cite{LamRueSed-20}. By looking at a source-destination link, two transduction steps are required \cite{KraRanHol-21}, as represented in Fig.~\ref{fig:03}:
\begin{itemize}
    \item[-] \textit{up-conversion}: converting the state of a superconducting qubit/ebit operating at microwave frequency $\omega_m$
into a degree of freedom  of an optical photon operating at frequency $\omega_o$.
    \item[-] \textit{down-conversion}: converting the state of a photonic qubit/ebit, operating at optical frequency $\omega_o$, into the state of a superconducting qubit/ebit, operating at microwave frequency $\omega_m$.
\end{itemize}
However, there exists a non-zero probability that either or both the conversions fail, with success-probability values strictly depending on the particulars of the hardware used for implementing the microwave-optical transduction. Among the different factors driving the success --  equivalently, the failure -- of a transduction attempt, which include added noise, conversion bandwidth, and mode-shape mismatch, the conversion efficiency\footnote{Formally defined in Sec.~\ref{sec:03.3}.} plays the role of the key hardware parameter. Low values of conversion efficiency, namely, high values of failure probabilities -- inevitably affect the performances of the transduction, and thus of the overall performances of the quantum communication link, as we will further explain in Sec.~\ref{sec:03.3} and Sec.~\ref{sec:03.4}.
Great efforts are being actively put on designing and improving the experimental devices. However, despite a huge progress has been made in the past decades, preserving the quantum states during conversions is still hard to reach with the state-of-the-art technology \cite{ZhoHanJia-22}.

In this regard, on one hand bulk optomechanical transducers achieved the highest conversion efficiency in literature about $50\%$ \cite{AndPetPur-14, Higginbotham2018}. On the other hand, bulk electro-optical transducers have achieved lower efficiency $10\%$ \cite{SahHeaRue-22}, but with much lower added noise ($\lesssim1$). This allowed for the demonstration of microwave-optical photon entanglement generation \cite{Sahu2023}, as we deeply discuss in Sec.~\ref{sec:04}. Integrated transducers \cite{Fan2018, Mirhosseini2020, Han2020, HolSinZhu-20, Xu2021LN, Hoenl2022, Jiang2023, Blesin2024, Meesala2024PRX} can offer more compact device footprint and high scalability, but the efficiency so far is limited to up to a few percent, mostly due to the lower power handling capabilities. 
Furthermore, integrating superconducting qubits within the transducer device represents a major challenge, as does the reliable readout of superconducting qubits embedded in the transduction process.

Fig.~\ref{fig:milestones} provides a summary of key literature highlighting the major advancements and milestones in microwave-optical (M-O) quantum transduction. The timeline represented includes the earliest and most significant demonstrations of microwave–optical transduction, highlighting the first relevant results in conversion efficiency and communication-oriented implementations.  Specifically, early advances include the realization of high-efficiency bulk-optomechanical transducer \cite{AndPetPur-14} and the first integrated electro-optic transducer \cite{FanZouCha-18}. Subsequent steps show the first integration between a superconducting qubit and an quantum transducer \cite{Mirhosseini2020}, followed by quantum transducer implementation achieving low added noise \cite{SahHeaRue-22}. More recent demonstrations include the first optical readout of a superconducting qubit \cite{DelMitBru-22} and the generation of entanglement between microwave and optical fields \cite{Sahu2023}. The works featured here are not meant to be exhaustive, but rather to illustrate representative milestones that have significantly shaped the evolution of the field.

\subsection{Direct Transduction of Information Carriers}
\label{sec:03.1}

DQT acting on an informational qubit is very far from being a trivial process, since it must be able to preserve the encoded quantum information \cite{KylRauWar-23}. To this aim, it is necessary that the quantum information is preserved through both the transmission on the quantum channel and during the up- and down-conversion processes. However, if the qubit is lost during the transmission due to the channel attenuation or is corrupted by noise, the associated quantum information cannot be recovered by a measurement process or by re-transmitting a copy of the original information, due to the quantum measurement postulate and the No-Cloning Theorem \cite{CacCalVan-20}. 

Accordingly, by taking into account both the impairments induced by the channel transmission and the failure probability of conversion processes, we conclude that DQT on informational qubits is not yet a viable strategy for the today technology, as analytically proved in \cite{DavCacCal-24, Tsa-10, Tsa-11}.



\begin{table}[t!]
    \centering
    \normalsize
    \renewcommand{\arraystretch}{0.9} 
    \begin{tabular}{cc}
        \textbf{Symbol} & \textbf{Hardware Parameter} \\ 
        \midrule
        \addlinespace[0.6ex]
         $\omega_m$  & microwave frequency   \\
         $\omega_o$  & optical frequency   \\
         $\omega_p$  & optical pump frequency   \\
         $\eta$ & conversion efficiency \\
        $C$  & cooperativity   \\
        $\zeta_x$  &  extraction ratio of mode $x$  \\
        $\kappa_x$  &  total dissipation rate of mode $x$\\
        $\kappa_{x,e}$  &  external coupling rates of mode $x$ \\
        $g$ & single-photon electro-coupling rate \\
        $n_p$ & pump photon number \\
        \bottomrule
    \end{tabular}
    \caption{Main transducer hardware parameters.}
    \label{tab:02}
    \hrulefill
\end{table}

\subsection{Direct Transduction of Entanglement Carriers}
\label{sec:03.2}

By considering the limitations of DQT on informational qubits, an alternative approach is to apply up- and down-conversions on the entanglement resource itself, i.e., on the ebits.
By adopting this strategy,the impact of noisy quantum transduction and noisy optical propagation shifts from quantum information to entanglement resource. Thus, its main advantage lies in the possibility of entanglement regeneration. 

Indeed, differently from informational qubits, entanglement -- being a communication resource rather than information -- is not constrained by the no-cloning theorem \cite{IllCalMan-22}. Thus, even if the ebit carrying quantum correlation is lost during the channel transmission or the transduction conversion fails, it can be regenerated without restrictions, until the conversion finally succeeds and the entanglement is correctly distributed between the remote nodes. Once the entanglement distribution is successful, an informational qubit can then be ``transmitted'' via quantum teleportation.
This allows to overcome the stringent requirements of DQT on quantum information.
The main differences between DQT on quantum information and on ebits are summarized in Table~\ref{tab:03}.

\subsection{Conversion efficiency}
\label{sec:03.3}

Generally, for implementing both up- and down- conversion of a quantum carrier, an input laser pump is required to facilitate the conversion of the photon associated to the qubit/ebit to be converted.

The conversion of the input photon into the output photon at the desired frequency can be performed through one or more intermediate steps, such as mechanics \cite{Tsa-10, Tsa-11, AndPetPur-14, Higginbotham2018, Mirhosseini2020, Han2020, Jiang2023} or magnonics \cite{Zhu2020Optica, Zhu2022PRApplied}, depending on the transducer hardware. Direct conversion (between microwave and optical frequencies) is instead realized through \textit{electro-optic transducers} \cite{Tsa-10, Tsa-11, HolSinZhu-20, FuXuLiu-21, Xu2021LN, SahHeaRue-22, Sahu2023}, which reduce device complexity and avoid intermediate noise sources. The trade-off though is the weaker nonlinearity compared with optomechanical schemes.
For the sake of clarity, in this paper we focus on electro-optic quantum transducers, but the developed theoretical analysis can be easily extended to different transduction hardware, by properly accounting for the particulars of the hardware parameters\footnote{See as instance the conversion efficiency (Eq. 6) in \cite{HanFuZou-21} for electro-optomechanical transducers.}.
Typically, an electro-optical quantum transducer consists of an optical cavity coupled with a microwave resonator. By (over) simplifying, an input laser pumped in the optical cavity initializes the Pockels effect, which enables a \textit{beam-splitter interaction} between optical and microwave signals \cite{Tsa-10, Tsa-11} providing a direct microwave-optical photon conversion. Specifically, the pump laser at frequency 

\begin{figure}[t!]
    \centering
    \includegraphics[width=1\columnwidth]{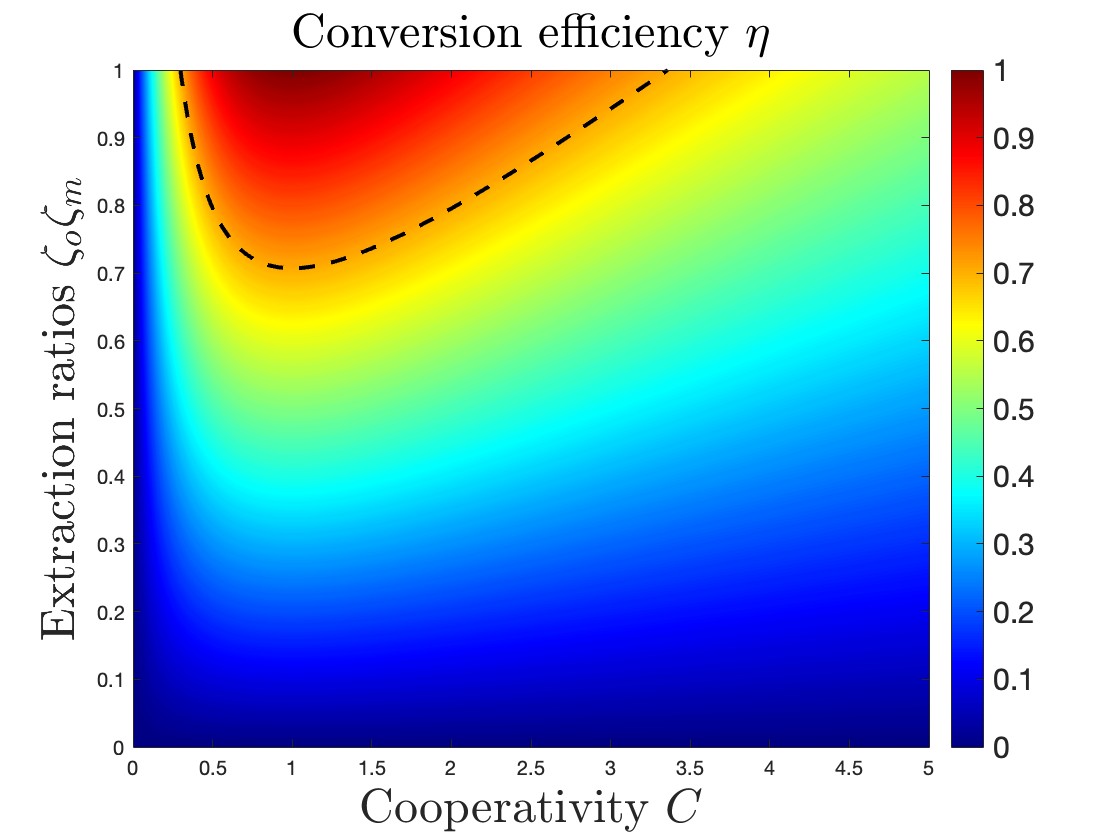}
    \caption{Conversion efficiency $\eta$ as a function of cooperativity $C$ and the product of extraction ratios $\zeta_o\zeta_m$.}
   \label{fig:conversion}
   \hrulefill
\end{figure}

$\omega_p=\omega_o-\omega_m$ interacts with an input photon at microwave frequency frequency $\omega_m$ (at optical frequency $\omega_o$) to produce an optical photon at $\omega_o$ (a microwave photon at $\omega_m$). 

The main parameter governing electro-optic transduction is the \textit{conversion efficiency} $\eta$, which denotes the probability of successful conversion. In the resolved-sideband limit where undesired amplification is negligible, the system is reciprocal and the efficiency is the same for up- and down-conversions\footnote{Although the added noise can in general differ between the two different conversion directions, depending on how the noise sources couple to the conversion process, as detailed in \cite{HanFuZou-21}.}. 
Under resonant conditions, the efficiency can be expressed as \cite{Tsa-11}:
    \begin{equation}
        \label{eq:04}
        \eta = 4 \zeta_o \zeta_m \frac{C}{|1+C|^2}.
    \end{equation}
In Eq. \eqref{eq:04}, $\zeta_x$ denotes the so-called extraction ratio of mode $x$\footnote{We denote optical mode with subscript $o$ whereas microwave mode with subscript $m$.}, given by the ratio between the external coupling rates $\kappa_{x,e}$ and the total dissipation rates $\kappa_{x}$, i.e., formally $\zeta_x=\frac{\kappa_{x,e}}{\kappa_x}$ \cite{HanFuZou-21}, and $C$ denotes \textit{cooperativity}, related to the interaction of microwave and optical field within the transducer, defined as \cite{Tsa-11, HanFuZou-21}:
  \begin{equation}
        \label{eq:cooperartivity}
       C = \frac{4 g_0^2 \, n_p}{\kappa_o\kappa_m}.
    \end{equation}

In Eq. \eqref{eq:cooperartivity}, $g_0$ denotes the single-photon electro-coupling rate and $n_p$ is the pump photon number. In the following texts, we use $\eta_{\uparrow}$ and $\eta_{\downarrow}$ to denote the efficiencies of up- and down- conversion, respectively. We choose this notation to enhance clarity of the paper, even though the formulation of the conversion efficiency, in terms of dependence on the transducer parameters, remains unchanged for both conversion directions. The main transducer parameters are summarized in Table~\ref{tab:02}.

By accounting for Eq.~\eqref{eq:04}, it follows that high conversion efficiency $\eta$ requires both cooperativity $C$ and extraction ratios $\zeta_x$ close to 1. This is clearly depicted in Fig.~\ref{fig:conversion}, which shows $\eta$ as a function of: i) $C$, and ii) the product of the extraction ratios $\zeta_o \zeta_m$. However, reaching high values for both these parameters is still an open and crucial challenge. Indeed, while there is a wide-scientific consensus in considering unitary values for $\zeta_x$ feasible to achieve in the near-future\footnote{Typical values assumed in theoretical studies are around $\zeta_x=0.9$ \cite{WuCuiFan-21}, whereas experimental values in the order of $0.1-0.2$ have already been measured \cite{FuXuLiu-21}.}, experimentally measured values for $C$ only recently exceeded  $0.3$ \cite{SahHeaRue-22} yet with an efficiency around $0.15$. 
Therefore, the cooperativity parameter constitutes the bottleneck of the transducer electro-optical efficiency.

\begin{remark}
    As pointed out in Sec.~\ref{sec:02}, quantum transduction is not just a merely frequency conversion process: many parameters challenge the interfacing between different hardware platforms such as photonic and superconducting.For instance, in order to achieve a good transduction, the physical modes of microwave and optical systems must be matched, which includes considerations of impedance, spatial overlap, and the temporal properties of the signals. This, can be captured by the electro-optic coupling coefficient $g_0$, as showed in \cite{RueSedCol-16}. 
    Accordingly, our choice of focusing on the conversion efficiency as the main characterizing parameter for quantum transduction has been key. In fact, this choice allowed us to abstract from the particulars of the specific technology underlying the transducer hardware. Consequently, the proposed analysis can be easily extended to different transducer hardware solutions, available in the state-of-the-art technology. And, remarkably, this choice allows us to track technological advancements by just adjusting a single parameter to incorporate the technological improvements.
\end{remark}

\subsection{Quantum Channel Capacity}
\label{sec:03.4}

In this section we discuss quantum transduction, and specifically DQT, with a perspective focused on the resulting \textit{quantum channel capacity}\footnote{The \textit{quantum capacity} denotes the maximum rate achievable through a quantum channel assuring a reliable transmission (i.e., the fidelity of the transmitted state is arbitrarily large), i) over asymptotically many uses of the channel, and ii) with the encoder allowed to generate entangled codewords \cite{KouCacCal-22, BennDiVSmo-97}.} of a communication system.

Since the conversion efficiency is the probability of having a successful conversion, low values of $\eta$ deeply affect the quantum channel capacity as well. More into details, having a non-zero quantum capacity imposes stringent conditions -- i.e., a minimum threshold value -- on $\eta$ \cite{WuCuiFan-21, ZhoHanJia-22}. And the actual value of the threshold depends on whether DQT occurs on the informational qubit or on the ebit.

More into details, the cascade of up-conversion, quantum
channel and down-conversion in a point-to-point communication link can be modelled as an overall
equivalent \textit{quantum erasure channel}, as deeply discussed in \cite{DavCacCal-24}.

In case of DQT on informational qubit, we consider the \textit{one-way quantum capacity} as the key parameter capturing the communication performances. Indeed, the one-way quantum capacity constitutes the default metric for a forward quantum channel alone unassisted by classical communication, as shown in \cite{BennDiVSmo-97}.

Accordingly, in \cite{DavCacCal-24} we showed that by safely ignoring the length effects of the fiber connecting the source and the destination, unitary one-way capacity requires unitary efficiency $\eta_\uparrow\eta_\downarrow=1$ in both the up- and the down- conversions \cite{DavCacCal-24}.Yet, such a value for the efficiency, as discussed in Sec.~\ref{sec:03.3}, exceeds current state-of-the-art technologies. In \cite{DavCacCal-24} we also showed that, by relaxing the hypothesis of unitary capacity, i.e., by requiring only a \textit{non-null} one-way quantum capacity, the up- and down efficiencies should satisfy the condition $\eta_\uparrow\eta_\downarrow > \frac{1} {2}$, i.e., a minimum threshold value that is still unachievable by state-of-the-art technology. In Fig.~\ref{fig:conversion}, the efficiency values enabling non-null one-way quantum capacity are highlighted with the dotted black curve.

When it comes to DQT acting on ebits, the cascade of up-conversion, fiber channel and down-conversion can be still modelled as an equivalent quantum erasure channel \cite{DavCacCal-24}. However, in this second scenario the quantum channel is assisted by two-way classical communication \cite{ZhoHanJia-22, KylRauWar-23}, and therefore the \textit{two-way quantum capacity} -- rather than the one-way capacity  -- should be considered as performance metric \cite{BennDiVSmo-97}.  Two-way quantum capacity can be greater than the correspondent one-way, and it is known to be positive for some channels for which the one-way capacity is zero \cite{BennDiVSmo-97}. An the rationale is that the ebits of the EPR pairs -- distributed for eventually teleporting the informational qubit to the destination -- can be regenerated and re-distributed in case of losses, without affecting the informational qubit. Thus, for assuring a \textit{non-null} two-way quantum capacity, in \cite{DavCacCal-24} we showed that the up- and down efficiencies should satisfy the condition $\eta_\uparrow\eta_\downarrow>0$, which is largely less stringent that the one imposed on DQT acting on informational qubit.

In a nutshell, the key advantage of DQT applied on ebits, with respect to DQT on informational qubits, is the largely less stringent requirement on the transducer hardware parameters for assuring a non-null quantum capacity. We will delve deeper on this in Sec.~\ref{sec:05.4}.
\begin{strip}    \tcbset{colback=white,colframe=black,colbacktitle=white,coltitle=black,colupper=black,fonttitle=\large\bfseries}
        \begin{tcolorbox}[oversize,title=\textsc{Entanglement}]
            \begin{multicols}{2}
                The most distinguish feature of quantum mechanics is entanglement, namely, a correlation with no counterpart in the classical world. Indeed, entanglement is considered the key communication resource for designing the Quantum Internet protocol stack \cite{KozWehVan-23, IllCalMan-22}, since it can be exploited to overcome the constraints induced by the \textit{no-cloning theorem} and the \textit{quantum measurement postulate} in the quantum communications domain. Whenever two qubits are entangled, the measurement of one of them instantaneously changes the state of the other, regardless of the distance separating the two qubits \cite{CacCalTaf-20}. Formally, given a state $\ket{\psi}$ of a composite quantum system, associated with the Hilbert space $V$, and a tensor decomposition of $V$, i.e., $V= V_0 \otimes V_1 \otimes \ldots \otimes V_{n-1}$, the state $\ket{\psi}$ is said to be \textit{separable or untangled} with respect to that decomposition, if it can be written as $\ket{\psi}= \ket{\psi_0} \otimes \ket{\psi_1} \otimes \ldots \otimes \ket{\psi_{n-1}}$, with $\ket{\psi_i} \in V_i$. Otherwise $\ket{\psi}$ is entangled with respect to that particular decomposition (but may be unentangled with other decompositions into subsystems) \cite{RiefPol-11, CacCalVan-20, IllCalMan-22}. Among the entangled states of two qubits, the \textit{Bell states}, called also EPR pairs, represents four maximally entangled 2-qubit states \cite{CacCalVan-20}:
                \begin{align}
                    \ket{\Phi^\pm} = \frac{1}{\sqrt{2}}(\ket{00}\pm\ket{11}) \\
                    \ket{\Psi^\pm} = \frac{1}{\sqrt{2}}(\ket{01}\pm\ket{10})
                \end{align}
                Maximally entangled states, as suggested by the name, provide the maximum amount of entanglement \cite{IllCalMan-22}. Despite the existence of various metrics for quantifying entanglement \cite{nielsen00}, there is widespread agreement in considering a pair of states as maximally entangled with respect to von Neumann entropy \cite{VedPleRip-97}.
            \end{multicols}
        \label{box:01}
    \end{tcolorbox}
\end{strip}

\section{Entanglement Generation Transduction (EGT)}
\label{sec:04}

\begin{figure}[t]
	\centering
	\begin{minipage}[c]{\linewidth}
		\centering
		\includegraphics[width=1\columnwidth]{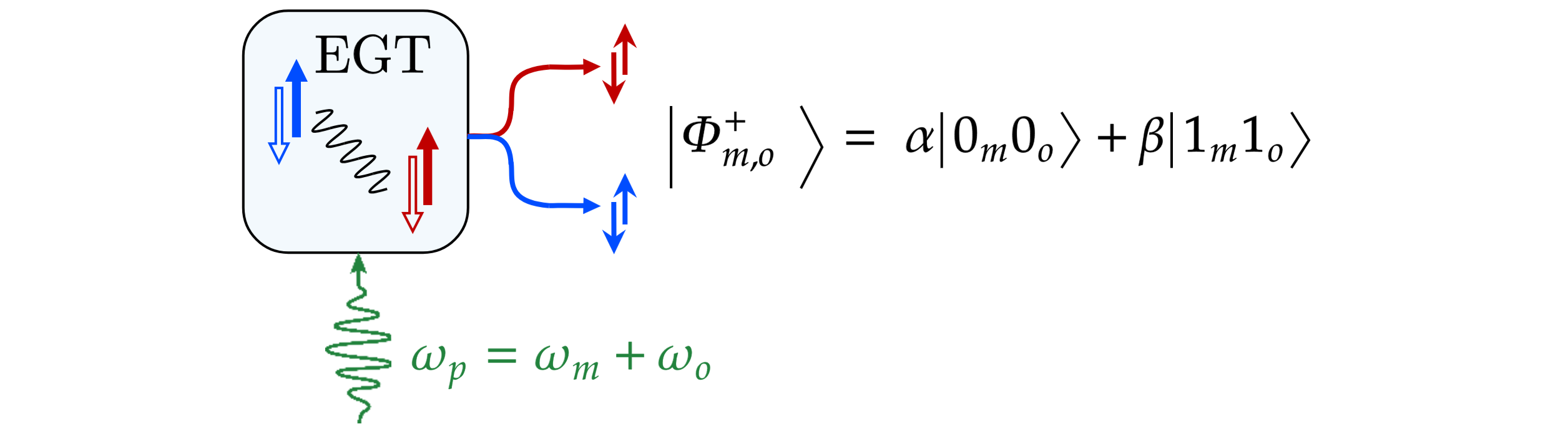}
		\subcaption{}
		\label{fig:x06.a}
	\end{minipage}
	\begin{minipage}[c]{\linewidth}
		\centering
		\includegraphics[width=1\columnwidth]{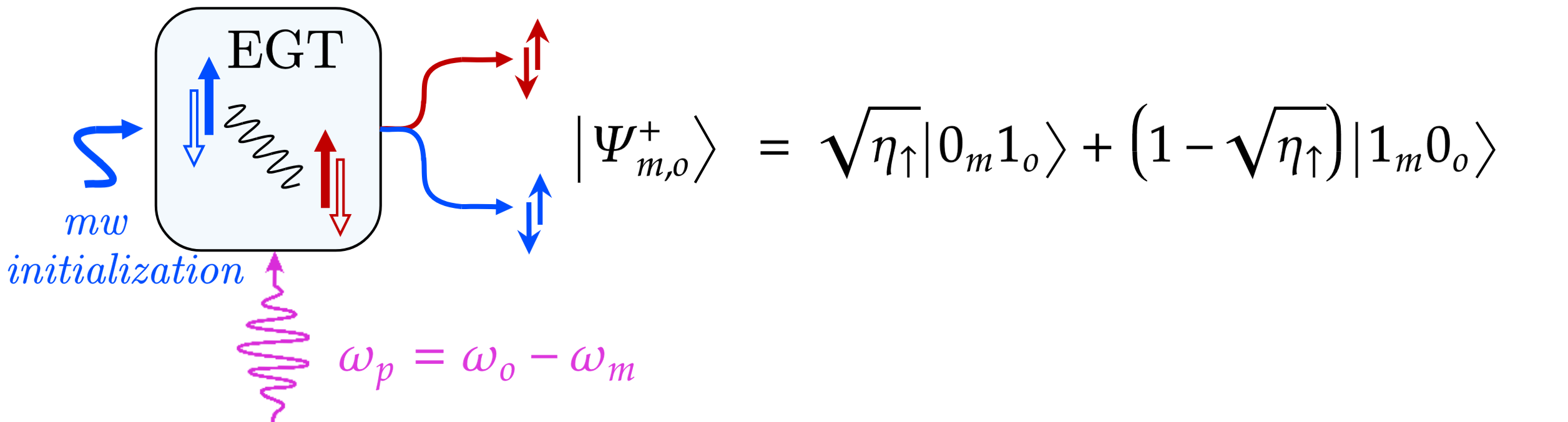}
		\subcaption{}
		\label{fig:x06.b}
	\end{minipage}
	\caption{Entanglement Generation Transduction (EGT) through (a) two-mode squeezing, and (b) beam-splitter interaction. Blue (red) ``up''-solid arrows represent the presence of a microwave (optical) photon, while blue (red) ``down''-empty arrows denote the absence of a microwave (optical) photon.}
	\label{fig:x06}
	\hrulefill
\end{figure}

As introduced in Sec.~\ref{sec:02}, a quantum transducer can generate \textit{hybrid} microwave-optical entanglement \cite{Sahu2023, Meesala2024PRX} through the so-called \textit{entanglement Generation Transduction} (EGT), through two different electro-optic interaction: either \textit{two-mode squeezing} or \textit{beam-splitter} interaction, 
as shown in Fig.~\ref{fig:x06} and discussed in the next sections.


\subsection{EGT through two-mode squeezing interaction}
\label{sec:04.1}

In the case of EGT via \textit{two-mode squeezing} interaction \cite{Loudon2000, Wal-83}, by exciting the transducer the input pump drives a spontaneous parametric down-conversion (SPDC), which generates entanglement between optical and microwave fields \cite{WuCuiFan-21, RueHeaBar-19, ZhoWanZhi-20, KraRanHol-21}. More into details, in presence of the three-wave-mixing nonlinearity (such as in electro-optic devices), a pump photon at $\omega_p$ will have certain probability to split into a microwave photon at $\omega_m$ and an optical photon at $\omega_o$, where $\omega_p=\omega_m+\omega_o$ satisfies energy conservation. To enhance the probability of such SPDC process, high-Q cavities or resonant modes are usually needed for both microwave and optics. This interaction can be described as a $(a^{\dagger}b+b^{\dagger}a)$ in the Hamiltonian, also called two-mode-squeezing. As a result, entangled photons can be generated whenever the quantum transducer is initialized
with no input microwave photon, as depicted in Fig.~\ref{fig:x06.a}, and the input pump frequency is set to the sum of the frequencies of the optical and microwave photons, i.e., $\omega_p=\omega_m+\omega_o$, (a.k.a., “blue detuning”). Ideally, the output state can be expressed with Fock state notation as\cite{ZhoWanZhi-20, KraRanHol-21, HauHjeAnt-24}:
\begin{align}
    \label{eq:05}
    \ket{\Phi_{m,o}} \approx \alpha \ket{0_{m}0_{o}}+\beta \ket{1_{m}1_{o}}
\end{align}
with the subscripts $(\cdot_m)$ and $(\cdot_o)$ denoting the photon domain, i.e., microwave or optical. Accordingly, in Eq.~\eqref{eq:05} the term $\ket{1_{m} 1_{o}}$ denotes the generation of both microwave and optical photons, and the term $\ket{0_{m}0_{o}}$ denotes no photon generation \cite{ZhoWanZou-20}. 
The coefficients $\alpha$ and $\beta$ depend on the hardware parameters as the effective squeezing factor \cite{ZhoWanZhi-20} and the cooperativity parameter $C$.

It is important to note that Eq.~\eqref{eq:05} assumes that the generation of higher order photon pairs is negligible, which is a good approximation in the low power regime. In general, the SPDC produces a two-mode squeezed vacuum that gives continuous-variable entanglement \cite{ZhoWanZou-20}.

\subsection{EGT through beam-splitter interaction}
\label{sec:04.2}

The above assumption of neglecting the generation of higher order photon pairs in EGT with two-mode squeezing interaction -- which is reasonable, as said, in the low power regime -- is satisfied without any restriction in case of EGT based on \textit{beam-splitter} interaction \cite{Gerry2023, FeaLou-87}. In this case, the transducer requires a specific initialization consisting of a microwave photon inside the cavity  \cite{KraRanHol-21} -- as schematically depicted in Fig.~\ref{fig:x06.b} -- and the input pump field is set to operate on a frequency that is the difference of the frequencies of the optical and microwave photons, i.e. $\omega_p=\omega_o-\omega_m$ (a.k.a. ,``red detuning'').
This leads to an entangled state in the form\cite{DavCacCal-2024}:
\begin{align}
    \label{eq:06}
    \ket{\Psi_{m,o}} = \sqrt{\eta_\uparrow}\ket{0_m 1_o}+(1-\sqrt{\eta_\uparrow})\ket{1_m 0_o},
\end{align}
where the term $\ket{0_{M} 1_{O}}$ denotes that the microwave photon of the initialization has been converted into an optical one, and the term $\ket{1_{M}0_{O}}$ denotes that the microwave photon was not converted. Specifically, if the transducer conversion efficiency is $50\%$ the state in Eq.~\eqref{eq:06} becomes \cite{DavCacCal-24}:
    \begin{align}
    \label{eq:EPR}
        \ket{\Psi_{m,o}} = \frac{1}{\sqrt{2}}(\ket{0_m 1_o}+\ket{1_m 0_o}),
    \end{align}
that constitutes a Bell State between different frequency domains.

\begin{remark}
      In this paper we only consider the effect of the conversion efficiency on the purity of the generated state, as in Eq.~\eqref{eq:06}, but noise source and other hardware parameters must be take into account to obtain the ideal state of Eq.~\eqref{eq:EPR}.
     For instance, a correct initialization of the microwave photon in the optical resonator can influence the purity of the resulting state. Or significant added noise can introduce additional terms in Eq.~\ref{eq:06}. As result, any hardware mismatch from the ideal setting would impact on the purity of the generated entangled pair, and the assumption of obtaining an EPR state in the form of Eq.~\eqref{eq:EPR} depends on a careful setting of the transduction hardware parameters \cite{KraRanHol-21}.
\end{remark}

It is important to note that the beam-splitter interaction exploited in EGT is the same interaction exploited in DQT for frequency conversions (up- and down-) presented in Sec.~\ref{sec:03}. Therefore, this interaction seems to pose hardware challenges similar to those of DQT, i.e. low values of $\eta$ severely limiting the communication performances of the system. This is not the case, and the main differences between the beam-splitter interaction exploited for DQT and EGT, respectively, lay in two key points. First, the photon to be converted in the DQT is either the informational qubit we aim to transmit or the ebit of the EPR pair we aim to distribute. Therefore, these quantum states have to be preserved in the conversion. Conversely, in the EGT, the entanglement is generated in the quantum-degree-of-freedom represented by the presence/absence of a photon -- namely, the so-called path-entanglement \cite{BotKokBra-00, HuvWilDow-08, MonVerCap-17} -- and thus the preservation of the quantum states is not of concern.

The second key point is that, while low values of the efficiency $\eta$ imply higher conversion-failure probability in the DQT (either for informational qubit or ebit), in EGT low values of the efficiency $\eta$ do not imply a failure of the process, as evident from Eq.~\eqref{eq:06}. Indeed, $\eta$ determines how much the generated entangled state deviates from being a maximally entangled one. According to this last consideration, the great advantage of the EGT over DQT is therefore related to the achievable hardware parameters in the state-of-the-art technology. In other words, while the quality of DQT is strictly related to high values of conversion efficiency, the EGT process can generate entanglement with values of $C$ that are reachable with current state-of-the-art technology.

The main differences between DQT and EGT are summarized in Table~\ref{tab:03}.

It is worthwhile to note that, in EGT with beam splitter interaction, the role played by the microwave initialization at hardware level is reminiscent of a basis change, since Eq.~\eqref{eq:05} and Eq.~\eqref{eq:06} are equivalent quantum states up-to a basis change. Indeed, the entangled state generated by the two-mode squeezing interaction is an entangled generated from $\ket{0_o0_m}$, while the entangled state resulting from a beam splitter-type interaction corresponds to a state generated from $\ket{0_o1_m}$. This consideration helps to understand the role of microwave initialization at the hardware level as effectively implementing a basis change. Moreover, it allows us to use interchangeably the two EGT interactions, since they produce LOCC-equivalent states, thus equivalent states from a communication perspective.

\subsection{Quantum Channel Capacity}
In case of EGT, the generated entanglement is exploited for quantum teleportation.  Therefore, as in case of DQT on the ebits, the quantum channel is assisted by two-way classical communication and the metric adopted for communication performances evaluation is the two-way quantum capacity. The condition for no-null two way quantum capacity discussed in Sec.~\ref{sec:03.4} holds also for communication link exploiting EGT.

\begin{strip}
    \tcbset{colback=white,colframe=black,colbacktitle=white,
             coltitle=black,colupper=black,fonttitle=\large\bfseries}
    \begin{tcolorbox}[oversize,
                      title=\textsc{Quantum teleportation},
                      label={Box:QuantumTeleportation},
                      breakable]
        \begin{multicols}{2}
           Quantum teleportation is a key communication protocol in the Quantum Internet, allowing the transmission of a qubit without the physical transfer of the particle encoding the qubit \cite{CacCalVan-20}.
           The protocol requires an EPR pair shared between the source and the destination and classical communication, as pictorially depicted in the figure within this box. Specifically, quantum teleportation performs a Bell state measurement (BSM) on both the informational qubit $\ket{\psi}$ -- encoding the information to be transmitted -- and on the ebit at the source side of a previously shared EPR pair. The output of the BSM, that can be regarded as a \textit{pre-processing}, is a pair of two classical bits, encoding the measurement results on the two qubits at the source side. These two classical bits are sent to the destination through a classical channel. Once received, the destination then performs a \textit{post-processing}, which consists in applying a unitary operation on the ebit at its side, accordingly to the measurement outcomes. The result is that the original quantum state $\ket{\psi}$ has been teleported within the entangled qubit at the destination. It is essential to note that the measurement process included in the BSM implies the destruction of both the original qubit and the ebit at the source. Therefore, a subsequent teleportation requires a new EPR generation and distribution process.

            \begin{tcolorbox}[colback=white,colframe=white]
                \begin{adjustbox}{width=1\linewidth}
		            \begin{tikzcd}
		                & \lstick{$\ket{\psi}$}\gategroup[wires=2,steps=10,style={dashed,rounded corners,inner xsep=20pt,inner ysep=5pt}, background, label style={label position=above, yshift=-0.0cm,}, background]{\sc Source} &\qw &\ctrl{1} \gategroup[wires=2,steps=4,style={dashed,rounded corners,fill=gray!20,inner xsep=20pt,inner ysep=5pt}, background, label style={label position=below, yshift=0.16cm,}, background]{\sc BSM} & \gate{H} & \qw & \meter{}& \cw & \cw & \cwbend{3}  \\
			            \lstick[wires=3]{$\ket{\Phi^\texttt{+}}$} &  & \qw & \targ{} & \qw & \qw &  \meter{}& \cw & \cwbend{2} & & & \\
			            & \\
			            & \gategroup[wires=1,steps=10,style={dashed,rounded corners,inner xsep=20pt,inner ysep=5pt},background,label style={label position=below,yshift=-0.44cm,}, background]{\sc Destination} & \qw & \qw & \qw & \qw & \qw &\qw &  \gate{X} & \gate{Z}  & \qw \rstick{\sc $\ket{\psi}$}\\
			            &  &  &  & &  & &  &  & &  &
		            \end{tikzcd}
                \end{adjustbox}
            \end{tcolorbox}
        \end{multicols}
    \end{tcolorbox}
\end{strip}
Different network archetypes that exploit EGT for quantum teleportation are presented and deeply discussed in the next section.


\section{Source-Destination Link Archetypes}
\label{sec:05}

\begin{figure*}
    \hfill
    \begin{subfigure}{\linewidth}
        \includegraphics[width=\linewidth]{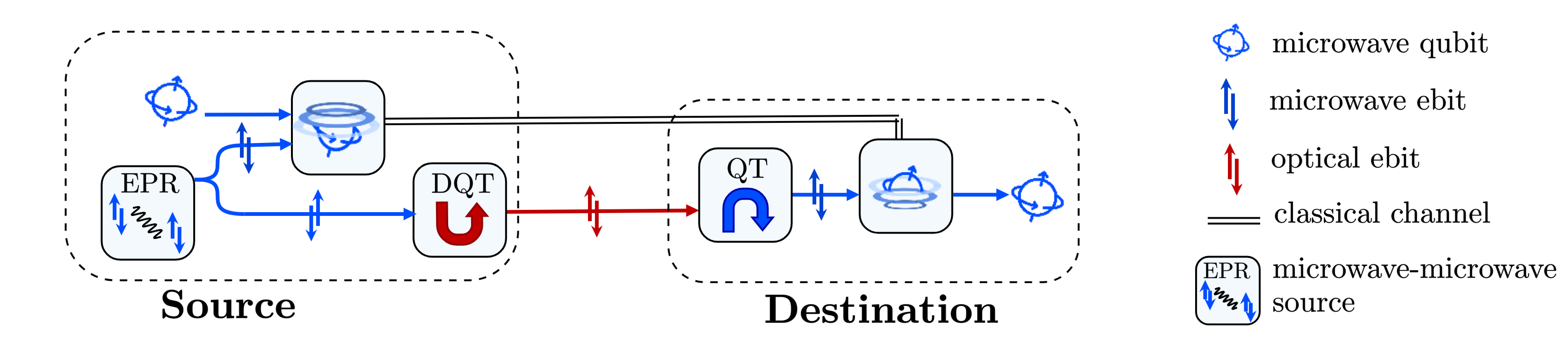}
        \caption{e-DQT}
        \label{fig:x07.1}
    \end{subfigure}
    \hfill
    \begin{subfigure}{\linewidth}
        \includegraphics[width=\linewidth]{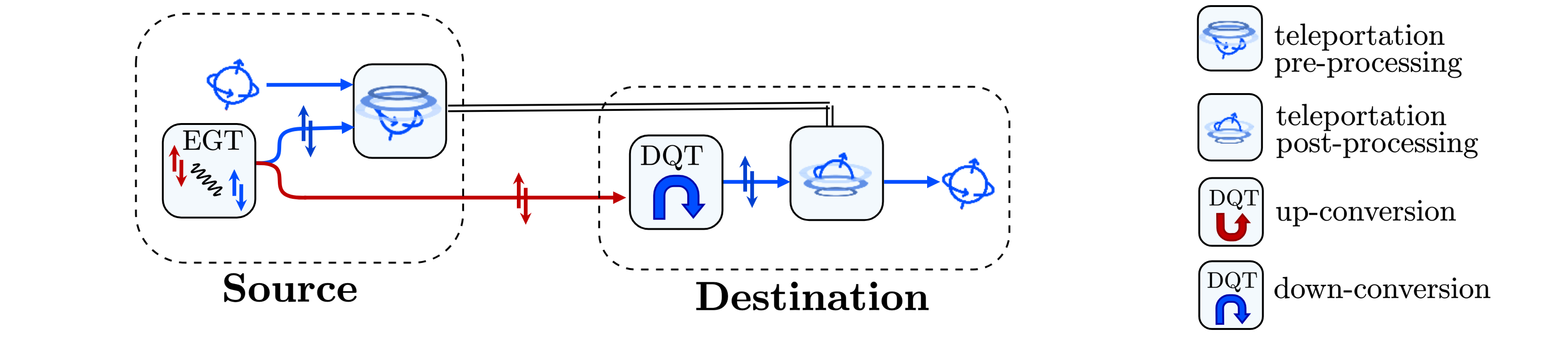}
        \caption{EGT Coupled with DQT}
        \label{fig:x07.2}
    \end{subfigure}
    \hfill
    \begin{subfigure}{\linewidth}
        \includegraphics[width=\linewidth]{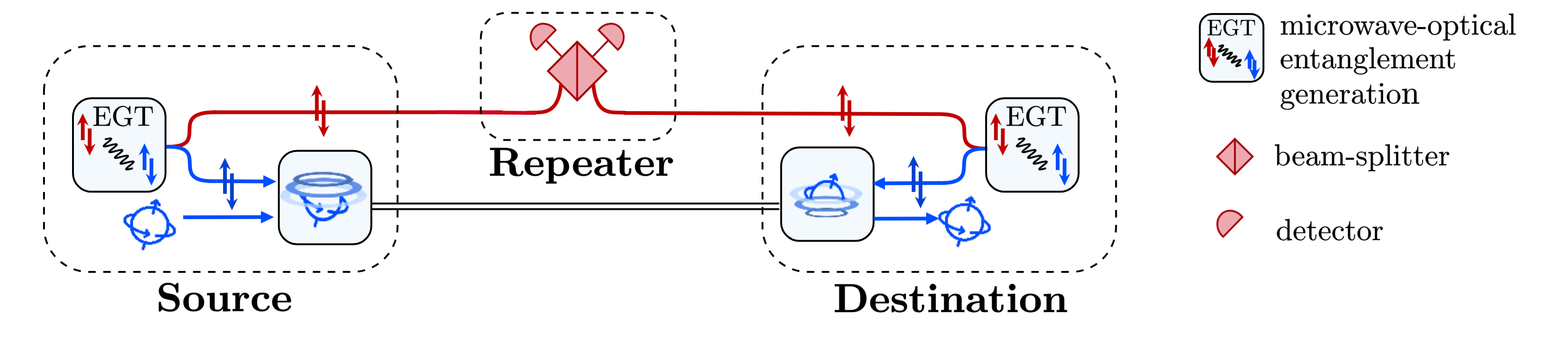}
        \caption{EGT Coupled with Swapping}
        \label{fig:x07.3}
    \end{subfigure}
    \caption{Source-Destination Link Archetypes}
    \hrulefill
    \label{fig:x07}
\end{figure*}

The analysis developed in the previous sections is not sufficient for grasping all the implications of QT on the design of a quantum network. Indeed, it is fundamental to configure QT into a specific network architecture for a more comprehensive overview. 
For the reasons highlighted in  Sec.~\ref{sec:03}, DQT on informational qubits is still beyond the state-of-the-art technologies, due to the stringent requirements in terms of high efficiency ($>50\%$) and low added noise ($\ll 1$). As a result, in the following, we focus only on the possibility to share quantum information among the network nodes via teleporting, thus transducing entanglement rather than informational qubits. Indeed, since quantum teleportation requires a pre-shared EPR pair between source and destination, we contextualize quantum transduction for entanglement generation and/or distribution.

Thus, in the next subsections we present different source-destination link archetypes leveraging both DQT on ebits and EGT. These archetypes, shown in Fig.~\ref{fig:x07}, recall the three well-known basic schemes for entanglement distribution on a link between two nodes, namely, ``at source'', ``at mid-point'' and ``at both end-points'' \cite{CacCalVan-20, KozWehVan-23}. Yet in the following we focus on the role of quantum transduction in each configuration, revealing where a quantum transducer should be placed as well as its functionality for each of the proposed archetypes \footnote{Although at the time of writing this manuscript the full implementation of the prosed network archetypes has not yet been achieved, each component of the experiment has been realized, such as the microwave readout from a transducer \cite{DelMitBru-22} and the photonic BSM \cite{BasRotSch-19, LuYanPan-09}.}.

\subsection{e-DQT}
\label{sec:05.1}

The first archetype we consider, referred\footnote{Let us highlight that in the previous section we used the acronym DQT referring to Direct Quantum Transduction, regardless of whether it is performed on the information carriers or entanglement carriers. Instead, we use the notation e-DQT to denote the specific source-destination link archetype described in this section, exploiting two DQTs of entanglement carriers.} in the following as e-DQT -- is a source-destination link where entanglement is first generated within the microwave domain and then it is distributed via two DQTs, as depicted in Fig.~\ref{fig:x07.1}. For the sake of simplicity, we assume that the entangled state to be distributed is an EPR pair. The state is locally generated at the source at microwave frequencies, which can be expressed in the Fock-state notation as \cite{YanFenZhi-20}:
\begin{equation}
    \label{eq:08}
    \ket{\Phi_{m,m}^{ss}}=\frac{1}{\sqrt{2}}(\ket{0_m^s0_m^s}+\ket{1_m^s1_m^s}),
\end{equation}
with the superscript $(\cdot^s)$ denotes the ``location'' of the photons, i.e., at the source.
The EPR is then distributed with a sequence of up- and down-conversions. Specifically, one microwave photon is up-converted at the source, sent over a fiber channel, and down-converted at the destination.
If both the conversions are successful, the resulting EPR state shared between the source and the destination is:
\begin{align}
    \label{eq:09}
    \ket{\Phi^{s,d}_{m,m}} = \frac{1}{\sqrt{2}}(\ket{0_{m}^{s}0_{m}^{d}}+\ket{1_{m}^{s}1_{m}^{d}}),
\end{align}
where the superscript $(\cdot^d)$ denoting the ``location'' of the photons at the destination.
Once the EPR is distributed the teleportation protocol can be performed.

As mentioned in Sec.~\ref{sec:03}, if one or both conversions fail, the entanglement generation and distribution process can be re-executed again until the distribution is successful.
Moreover, it is important to empathize that the informational qubit is not involved in the transduction process, but only in the local operations and classical communications (LOCC) \cite{CacCalTaf-20} required by the quantum teleportation protocol. This, in turn, implies that any failure of the DQT process do not impact the quantum state to be transmitted.

In Sec.~\ref{sec:05.5}, we generalize this archetype by removing the hypothesis of entanglement generated at the source.

\subsection{EGT Coupled with DQT}
\label{sec:05.2}

Here we present the second archetype, based on EGT and depicted in Fig.~\ref{fig:x07.2}. This archetype allows to reduce the number of direct conversions for distributing entanglement between source and destination with respect to the e-DQT archetype.

Specifically, a transducer located at the source generates hybrid entanglement, by exploiting one of the two physical interactions described in Sec.~\ref{sec:04.1} and Sec.~\ref{sec:04.2}, respectively. For the sake of simplicity, in the following, by assuming a beam spitter interaction and an ideal conversion efficiency of $\eta=50\%$, a Fock state in the form of Eq.~\eqref{eq:EPR} is generated.
The optical photon of the generated entangled pair is then transmitted to the destination through an optical fiber and down-converted to the microwave domain therein. The resulting state shared between the source and the destination can be expressed as follows:
\begin{equation}
    \label{eq:10}
    \ket{\Psi_{m,m}^{s,d}}=\frac{1}{\sqrt{2}}(\ket{0_m^s1_m^d}+\ket{1_m^s0_m^d}).
\end{equation}

In this archetype, the entanglement distribution process requires that the destination is equipped with a quantum transducer capable of down-converting (from optical to microwave domain) one ebit of the generated hybrid entangled state.  Thus, this archetype still suffers from the inefficiency of direct quantum transduction -- although limited to a single conversion (optical to microwave) rather than both up- and down-conversions.

\afterpage{
\begin{strip}   \tcbset{colback=white,colframe=black,colbacktitle=white,coltitle=black,colupper=black,fonttitle=\large\bfseries}
    \begin{tcolorbox}[oversize,title=\textsc{To go deeper: heralding entanglement}]
    \begin{multicols}{2}
           In \textit{EGT Coupled with Swapping}, entanglement heralding is assumed to be performed via photon-resolved detectors (PNRDs), namely, via detectors able to count the individual photons \cite{ProLukRa-20, CheZhoWan-22, KonZhaLiu-25}.
           With this setup, every detector click heralds a successful EPR distribution between the two remote nodes.
            However, when this hardware requirement cannot be satisfied, (cheaper) single-photon detectors (SPDs) can be used. SPDs are not able to distinguish whether a click is due to one or two temporally-coincident photons \cite{Had-09, EisFanMig-11, ZhaItzZbi-15}. Therefore, in this case, only a fraction of clicks corresponds to a distributed entanglement among distant quantum processors \cite{DavCacCal-24, DavZhaChu-24}.
            In other words, some detectors clicks do not reveal the presence of entanglement, but constitute dark counts.
            In addition and regardless from the adoption of either PNRDs or SPDs, the heralded clicks have to be weighted by the detector efficiency \cite{ProLukRa-20}, since a fraction of optical photons remains undetected due to non-unitary detection efficiency \cite{DavCacCal-24, DavZhaChu-24}. Therefore, the choice of the detector hardware impacts on the heralding capability. 
            This has been preliminarily analyzed in \cite{DavZhaChu-24}, by performing the first assessment via SeQUeNCe network simulator \cite{DavZhaChu-24} of the communication performances of quantum transduction in the archetype \textit{EGT Coupled with Swapping}.
            Although there is not yet an experimental validation of this network archetype, entanglement distribution between remote no-interaction systems have been experimental demonstrated with qubit platforms that do not require transduction, such as quantum dots \cite{BasRotSch-19} or ion traps \cite{LuYanPan-09}. Furthermore, the impact of the the choice of detector at the BSM node has been experimentally verified in \cite{JinTakTak-15, DoeSmiRos-17}.
            
            Furthermore, the heralding can also be affected by the type of interaction exploited within the EGT, namely, two-mode squeezing beam splitter vs beam splitter interaction described in Sec.~\ref{sec:04.1} and in Sec.~\ref{sec:04.2}, respectively.
            Indeed, the SPDC of the two-mode squeezing interaction can also generate more than one photon in the optical domain \cite{KraRanHol-21, LiYuaChan-25}.
            This implies that, beyond the case of $\ket{0^s_{m}}\ket{0^d_{m}}$, a detector click may erroneously herald a multi-photon state in the form $\ket{n^s_{m}}\ket{k^d_{m}}$, with $n,k\in \mathbb{N}$, as distributed entanglement. Consequently, exploiting two-mode squeezing interaction with SPDs as detectors further increases the occurrence of dark counts in entangled states. The issue of generating a multi-level system can be solved by exploiting beam splitter interaction with microwave initialization \cite{KraRanHol-21, Han-25}.
        \end{multicols}
        \label{box: EGT Coupled with Swapping}
    \end{tcolorbox}
\end{strip}
}

\subsection{EGT Coupled with Swapping}
\label{sec:05.3}

The third archetype is referred to as \textit{EGT Coupled with Swapping}\footnote{\textit{Entanglement swapping} \cite{BrieDurCir-98} is a strategy that extends the entanglement distribution distance. The reader may refer to the vast literature on the subject.} and it is depicted in Fig.~\ref{fig:x07.3}. 

Specifically, with two EGTs at the source and destination side, two hybrid EPR states in the form of Eq.~\eqref{eq:EPR} are generated. Accordingly, the entanglement generation occurs ``at both end- points'' rather than at ``source'' \cite{CacCalVan-20,KozWehVan-23}.

The optical ebits of both the generated entangled states are then transmitted through optical fibers, by reaching to a beam splitter followed by two detectors. The overall setup is configured so to be unable to distinguishing the \textit{which-path} information \cite{KraRanHol-21, PakZanTav-17,ZeuSorTay-20}. A click of one of the two detectors denotes the presence of an optical photon. However, due to the path-erasure -- i.e., the impossibility of knowing whether the optical photon responsible for the detector-click has been generated at the source or at the destination -- it is impossible to distinguish where the entanglement generation process has taken place (namely, whether at the source or at the destination), and thus it is impossible to distinguish whether a microwave photon is present at source or at destination. This results into the generation of another form of \textit{path-entanglement} \cite{MontVivCap-15} between the microwave photons at the source and at the destination, in the form of Eq. \eqref{eq:10}. Basically, the famous Duan–Lukin–Cirac–Zoller protocol is performed \cite{DuaLukCir-01}. Thus, the overall effect of beam splitter and detectors implements an entanglement swapping, projecting the received optical photons into a Bell state.

The distributed microwave-microwave entanglement can now be exploited to transmit quantum information through quantum teleportation. 

\begin{figure*}[ht!]
   \subfloat[e-DQT 
   \label{fig:x08.1}]{%
      \includegraphics[ width=0.33\textwidth]{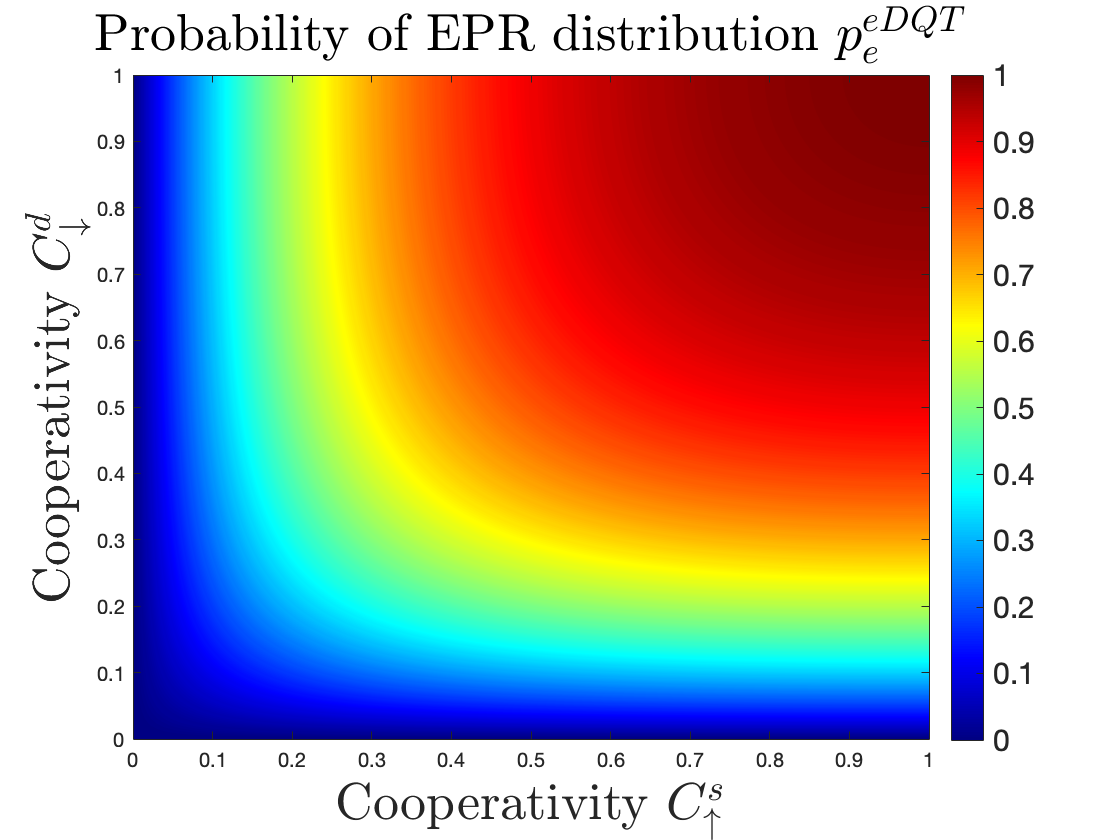}}
\hspace{\fill}
   \subfloat[EGT Coupled with DQT 
   \label{fig:x08.2} ]{%
      \includegraphics[ width=0.33\textwidth]{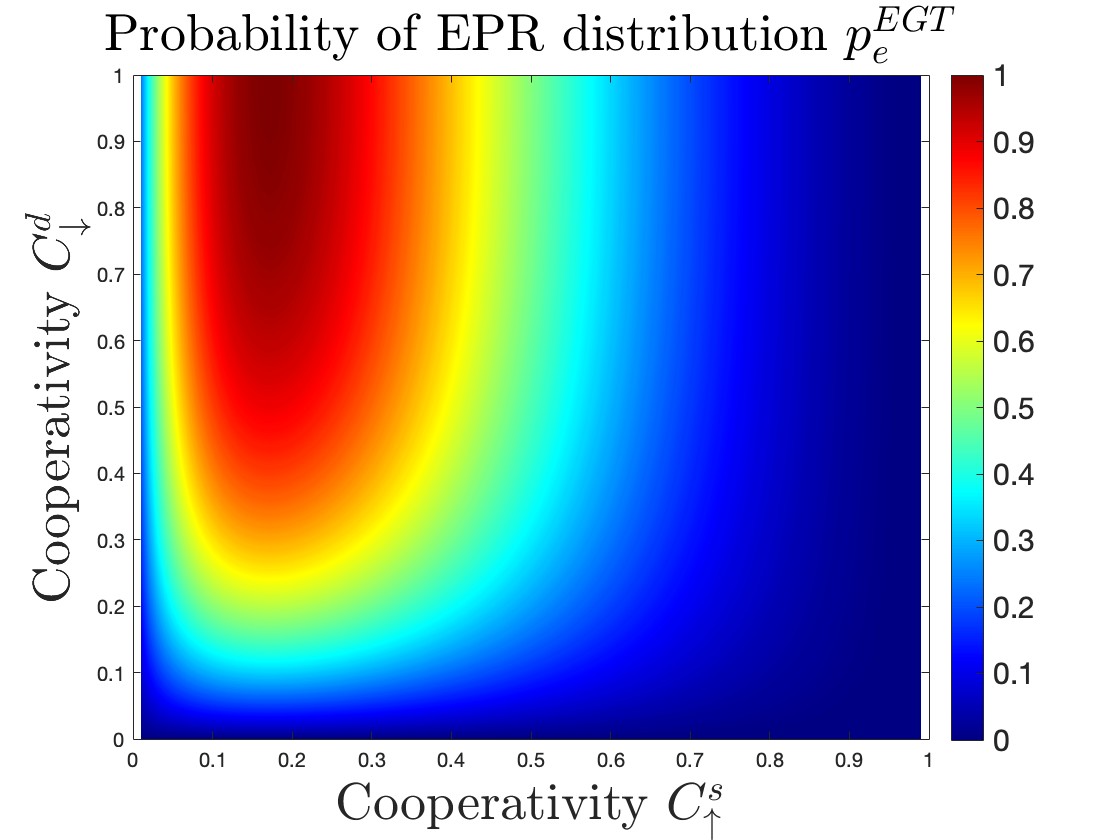}}
\hspace{\fill}
   \subfloat[EGT Coupled with Swapping 
   \label{fig:x08.3}]{%
      \includegraphics[ width=0.33\textwidth]{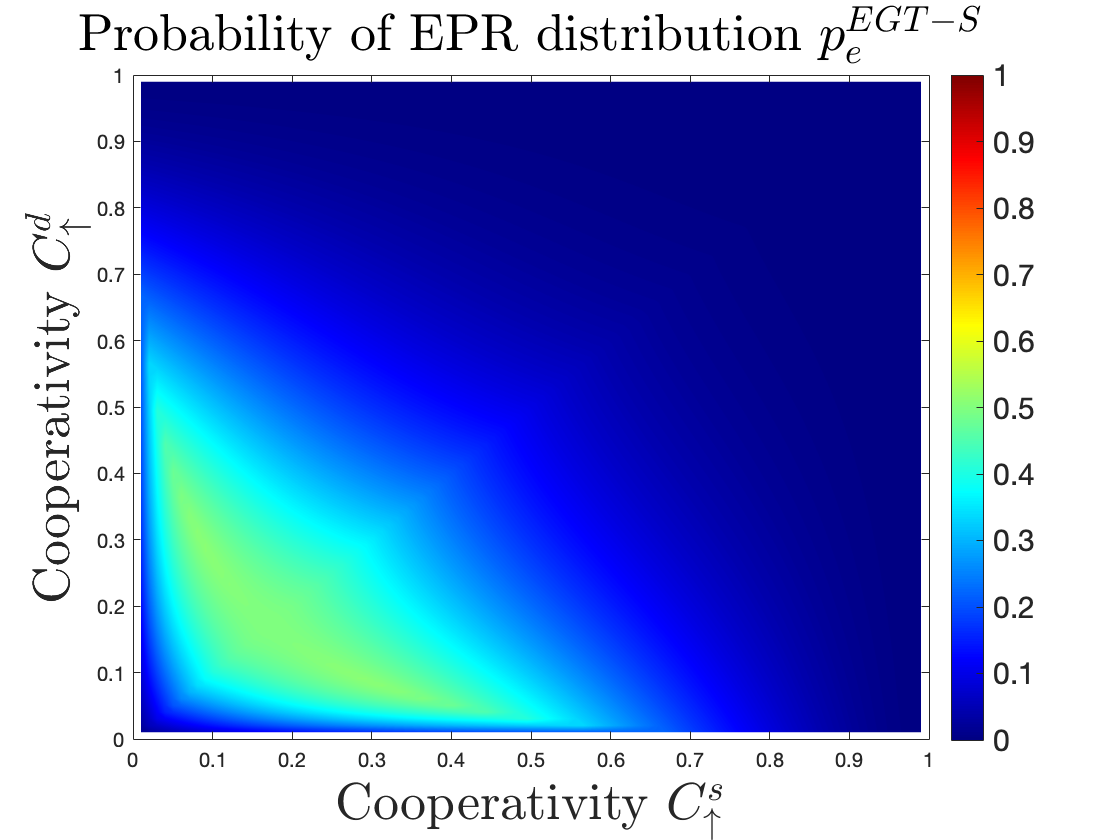}}\\
\caption{Probability of EPR distribution $p_e$ as a function of cooperativitie $C^s_\uparrow$ and $C^d_\downarrow$ for different Source-Destination Link archetypes}.
    \label{fig:x08}
\end{figure*}

\begin{figure}[t!]
    \centering
    \includegraphics[width=1\columnwidth]{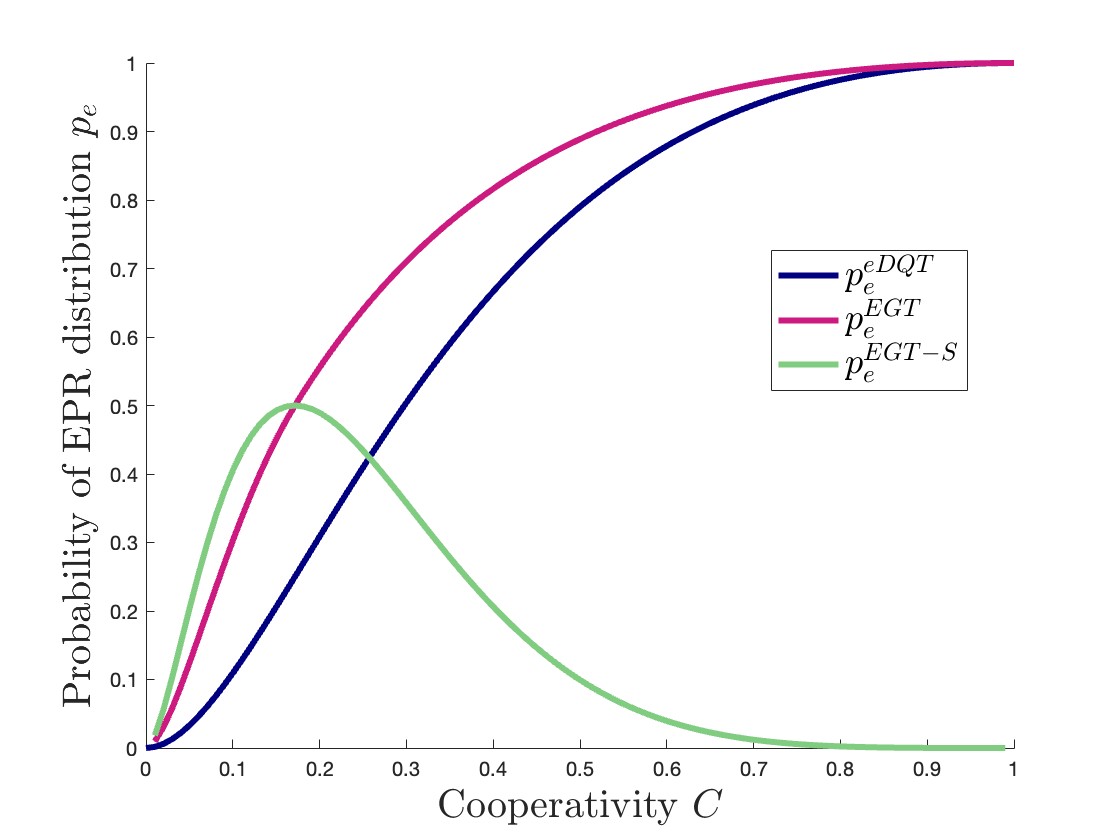}
    \caption{Performance comparison of the different transduction strategies: probability of EPR distribution $p_e$ as a function of cooperativity $C$, where $C=C^s_\uparrow=C^d_\downarrow$ for e-DQT and EGT Coupled with EGT and $C=C^s_\uparrow=C^d_\uparrow$ for \textit{EGT Coupled with Swapping}.}
    \label{fig:x09}
   \hrulefill
\end{figure}

It is worthwhile to highlight that a detector click does not always correspond to an EPR shared between source and destination. Indeed, while the purity of the hybrid generated EPR depends exclusively on the transducer hardware, as discussed in Sec.~\ref{sec:04.2}, the purity of heralded microwave EPR expressed in Eq. \eqref{eq:10} depends also on the characteristics of the repeater node, i.e. on the beam splitter and optical detectors characteristics. Specifically, there exists the possibility that more than one photon reach the repeater node. In fact, when both the transducers generate optical photons, only one detector click is triggered due to path erasure. In such a case, a detector click corresponds to the presence of two microwave photons, one at the source and one at the destination. Hence, the state shared between the remote nodes is $\ket{0^s_{m}}\ket{0^d_{m}}$, which is definitely not an entangled state as in Eq. \eqref{eq:10}.
However, if we reasonably assume the availability of photon-number-resolved detectors (PNRD), then it is possible to distinguish the event of receiving two optical photons -- one for each transducer in each link -- from the event where only one optical photon is received. And the double-photon event can be discarded in favour of a new distribution attempt.

It must be acknowledged that the key advantage of the archetype \textit{EGT Coupled with Swapping} lies in the possibility of heralding entanglement
via off-the-shelf hardware – i.e., via PNRD. Specifically, a
detector click for each transduction attempt constitutes an indicator for identifying the generation of entanglement between the source and the destination, without destroying it. 

In the previous two archetypes, namely, e-DQT and \textit{EGT coupled with DQT}, there exists the possibility of heralding entanglement by exploiting an additional degree of freedom, different from the one for entanglement encoding.  For instance, time-bin entanglement of microwave photons have been experimentally demonstrated \cite{KurPechRoy-19}. Also time bin of hybrid entanglement generated within a traducer (EGT) have been measured \cite{ZhoWanZou-20, Meesala2024PRX}. These examples not only exploit two different degree of freedom (one for entanglement generation and the other for entanglement heralding), but also imply the introduction of additional hardware setups for the heralding. On the contrary, in \textit{EGT Coupled with Swapping}, the entanglement heralding is embedded in the setup itself and, for this reason, it is not necessary to exploit other degree of freedom and/or to introduce any additional heralding
equipment.

\subsection{Archetypes Comparison}
\label{sec:05.4}

Stemming from the above described archetypes, here we conduct a performance comparison among them, by analysing some key communication metrics as functions of the main transducer-hardware parameters. 
Specifically, our analysis bridges the gap between different hardware platforms for quantum transducers, such as electro-optic and opto-mechanical systems. Indeed, as mentioned in Sec.~\ref{sec:03.3}, by focusing on common characteristics shared by all transducer implementations, such as conversion efficiency, we establish communication metrics that are agnostic to specific hardware configuration. 
The key metric we consider is the probability $p_e$ of successfully distributing an EPR from the source to the destination. For the theoretical details about the closed-form expressions of the probability of EPR distribution we refer the reader to the Appendices in \cite{DavCacCal-24}.

For the e-DQT archetype, $p_e^{eDQT}$, as a function of transducer parameters, can be expressed as follows \cite{DavCacCal-24}:
\begin{equation}
    \label{eq:12}p_e^{eDQT}=\eta_\uparrow^s\eta_\downarrow^de^{-\frac{l_{s,d}}{L_0}},
\end{equation}
where the superscripts $(\cdot^s)$ and destination $(\cdot^d)$ denote the ``location'' of the transducer process, i.e., at source and at destination side.
In Eq.~\eqref{eq:12} the term $e^{-\frac{l_{s,d}}{L_0}}$ takes into account the fiber losses, with $l_{s,d}$ denoting the length of the fiber link between source and destination and $L_0$ denoting the attenuation length of the fiber. As for today, commercial optical fibers feature an attenuation lower than $1$db/km. As instance, optical photons with wavelength in around $1550$nm -- i.e., with wavelength within the DWDM ITU C band -- experience an attenuation of roughly $0.2$dB/km, which corresponds to $L_0=22$ km 
\cite{SanChrder-11}.

It is evident that in case of noiseless quantum teleportation, i.e, under the hypothesis of noise-free LOCC, $p_e$ can be seen as the probability of successful transmission of quantum information as well.
Therefore, the probabilities of successfully transmitting quantum information and successfully distributing entanglement become equivalent. In other words,  the two-way quantum capacity coincides with $p_e$.

\begin{table*}[t]
    \centering
    \includegraphics[width=\textwidth]{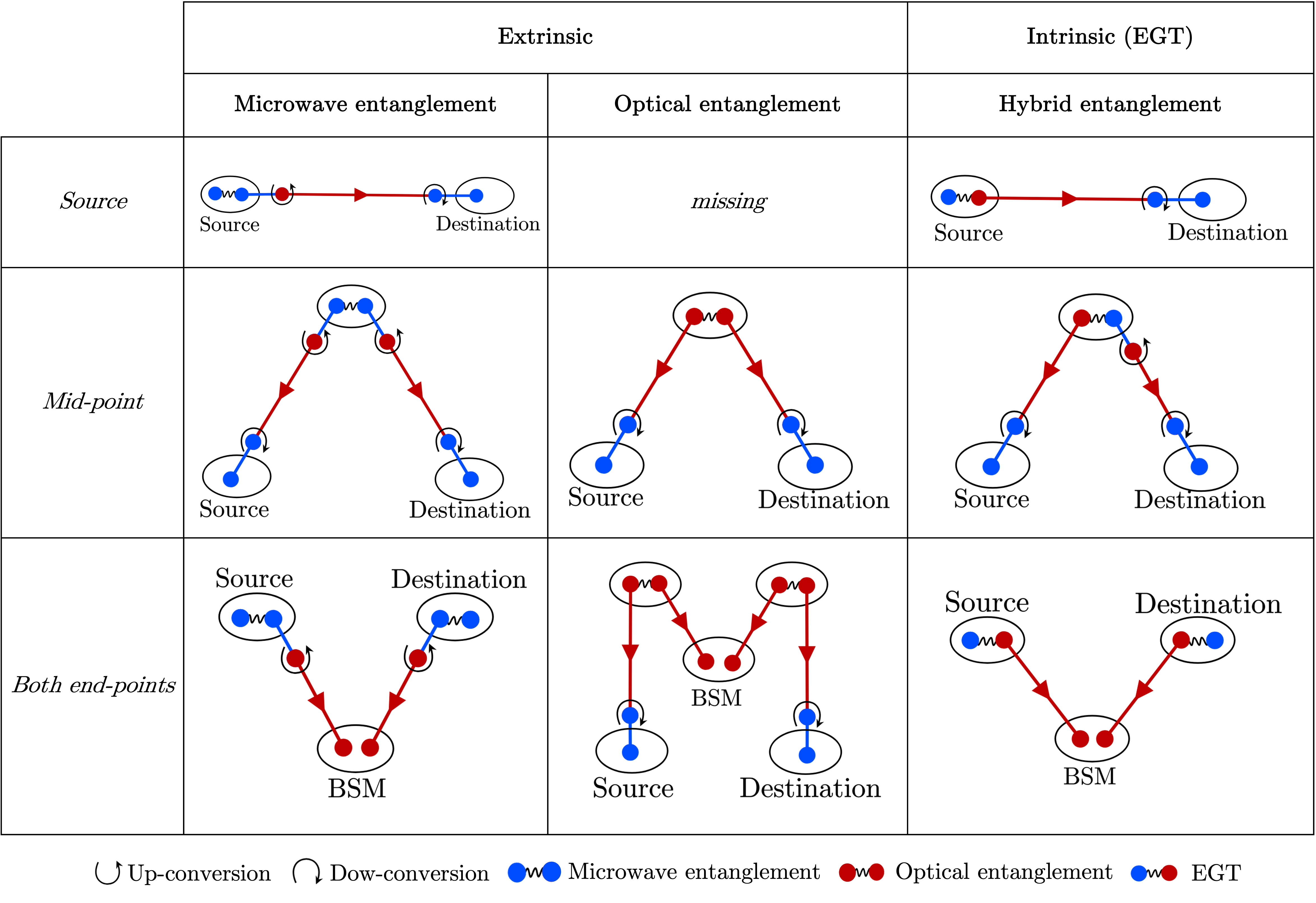}
    \caption{Source-destination link Archetypes. Blue (red) dots denote the microwave (optical) photons.}
    \label{tab:04}
\end{table*}

For the \textit{EGT Coupled with DQT} archetype, the probability $p_e^{\text{\rm EGT}}$ of successfully distributing an EPR pair is:
\begin{equation}
    \label{eq:pEQT}
    p_e^{\text{\rm EGT}} = S \big( \eta^s_{\uparrow} \big) \eta^d_{\downarrow}e^{-\frac{l_{s,d}}{L_0}}
\end{equation}
where $S(\cdot)$\footnote{With a small abuse of notation, we have indicated in the argument of the Von Neuman entropy the eigenvalue determining its value rather than -- as usually done -- the density matrix on which the entropy is evaluated. For further details please refer to \cite{DavCacCal-24}.} denotes the Von Neuman entropy given by:
\begin{equation}
    \label{eq:entropy}
    S(\eta^s_{\uparrow}) = - \eta^s_{\uparrow} \log_2(\eta^s_{\uparrow}) - (1-\eta^s_{\uparrow})\log_2 (1-\eta^s_{\uparrow}).
\end{equation}

Finally, for the \textit{EGT Coupled with Swapping} archetype, the probability $p_e^{\text{\rm EGT-S}}$ of successfully distributing an EPR  is given by \cite{DavCacCal-24}:
    \begin{equation}
        \label{eq:pEQTswap}
       p_e^{\text{\rm EGT-S}} = S(\tilde{\eta}_\uparrow)\big[ \eta_\uparrow^s(1-\eta_\uparrow^d) + \eta_\uparrow^d(1-\eta_\uparrow^s)\big]e^{-\frac{l_{s,d}}{2L_0}},
    \end{equation}
where $\tilde{\eta}_\uparrow$ denotes the efficiency between $\eta_\uparrow^s$ and $\eta_\uparrow^d$ that minimizes $S(\cdot)$.  

Fig.~\ref{fig:x08} shows the probability $p_e$  of successful EPR distribution for the three considered archetypes, as function of the cooperativity $C$ (namely, as discussed in Sec.~\ref{sec:03.3}, the main hardware parameter limiting the transducer performances). We follow for $C$ the same notation adopted for $\eta$, thus we indicate with the subscripts $(\cdot_\uparrow)$ and $(\cdot_\downarrow)$ the direction of the conversion (up or down) and with the superscripts $(\cdot^s)$ and $(\cdot^d)$ the location of the transduction (source and destination). With reference to the e-DQT archetype, the presence of two DQT processes requires unitary values for both $C^s_\uparrow$ and $C^d_\downarrow$ in order to obtain $p_e=1$ (equivalently unitary two-way quantum capacity), as shown in Fig.~\ref{fig:x08.1}. This restriction can be relaxed in the case of \textit{EGT Coupled with DQT}. In fact, in this archetype, to obtain a unitary $p_e$, the value of $C$ strictly equal to 1 is required only for the transducer at the destination ($C^d_\downarrow$=1), responsible for the direct conversion, while $C^s_\uparrow$ (driving the entanglement generation at the source) must satisfy $C^s_\uparrow \approx 3-2\sqrt{2}$, as shown in Fig.~\ref{fig:x08.2}.

Finally, in \textit{EGT Coupled with Swapping}, no direct conversion is required. Therefore, the maximum amount of $p_e$ can be achieved with $C^s_\uparrow=C^s_\uparrow\approx 3-2\sqrt{2}$, as shown Fig.~\ref{fig:x08.3}. Thus, this archetype determines an improvement in terms of minimum cooperativity that allows a non-zero entanglement distribution probability. But, this comes at the cost of a probability $p_e$ that never reaches $1$, which, in turn, implies that the two-way quantum capacity does not reach one as well.

To further highlight the comparison between the performances of the different proposed archetypes, Fig.~\ref{fig:x09} presents the three $p_e$ as function of $C$ within the same plot\footnote{Here, for \textit{EGT Coupled with DQT}, $C$ is $min(C, 3-2\sqrt{2})$.}, Specifically, in Fig.~\ref{fig:x09} we assume $C^s_\uparrow=C^d_\downarrow=C$ for both e-DQT and \textit{EGT Coupled with DQT}, while we assume $C^s_\uparrow=C^d_\uparrow=C$ is considered for \textit{EGT Coupled with Swapping}.

\begin{remark}
    The performance comparison in terms of EPR distribution probability $p_e$ in Fig.~\ref{fig:x08} and Fig.~\ref{fig:x09} maps into an equivalent capacity assessment. Indeed, the probabilities of EPR distribution constitute a capacity upper-bounds of the proposed network archetypes, and we refer the reader to \cite{DavCacCal-2024} for an in-depth technical discussion about this aspect.
\end{remark}

It is evident that, in archetypes exploiting transducers generating hybrid entanglement, it is possible to relax the constraints on the required hardware parameters. Conversely, the presence of a DQT acts as a bottleneck, limiting the overall performances of the system.

\subsection{Additional Source-Destination Link Archetypes}
\label{sec:05.5}

The analysis developed in the previous subsections is not exhaustive. Indeed, additional source-destination link archetypes can be considered, accordingly to i) the entanglement generation ``location" and ii) the entanglement type.

More in detail, as we mentioned at the beginning of this section, with reference to the ``location'' the entanglement generator can be either ``at source'', ``at mid-point'' or ``at both end-points'' \cite{CacCalVan-20,KozWehVan-23}. Instead, regarding the type, it is possible to distinguish archetypes depending on the entanglement resource is generated within (namely, EGT described in Sec.~\ref{sec:04}) or outside the transducer, through an external entanglement source. In literature, it is common to refer to these two types of entanglement generation as \textit{intrinsic} and \textit{extrinsic} process \cite{KylRauWar-23}, respectively. 

Table~\ref{tab:04} summarizes the possible source-destination link archetypes, accordingly to the ``location'', frequencies and type of entanglement, including but not limiting to the archetypes discussed in the previous subsections. Teleportation process for quantum information transmission is not explicitly depicted to make the figures clearer and, as done before, blue and red symbols refer to ebit at microwave and optical frequencies, respectively. By observing the various possibilities, we can draw some considerations.

First, it is worth to highlight that, as the number of DQT processes increases, the risk of losing the entanglement increases accordingly. Indeed, any processing can only worsen entanglement. This directly implies that extrinsic entanglement generation at ``mid-point" requires at least two DQT processes that can scale up to four in the case of entanglement generated at microwave frequencies.

Additionally, we observe that it is meaningless to treat the generation of extrinsic optical entanglement at ``source", since we are considering the network scenario where source and destination are superconducting nodes. Furthermore, the generation of extrinsic optical entanglement at ``both end-points'' introduces a communication archetype that differs from those introduced in \cite{CacCalVan-20, KozWehVan-23}. Indeed, the ``both points'' referred to are not source and destination nodes, but two external optical nodes, whose function is just to generate entanglement.
\begin{figure*}[h!]
    \centerline{\includegraphics[width=1\textwidth]{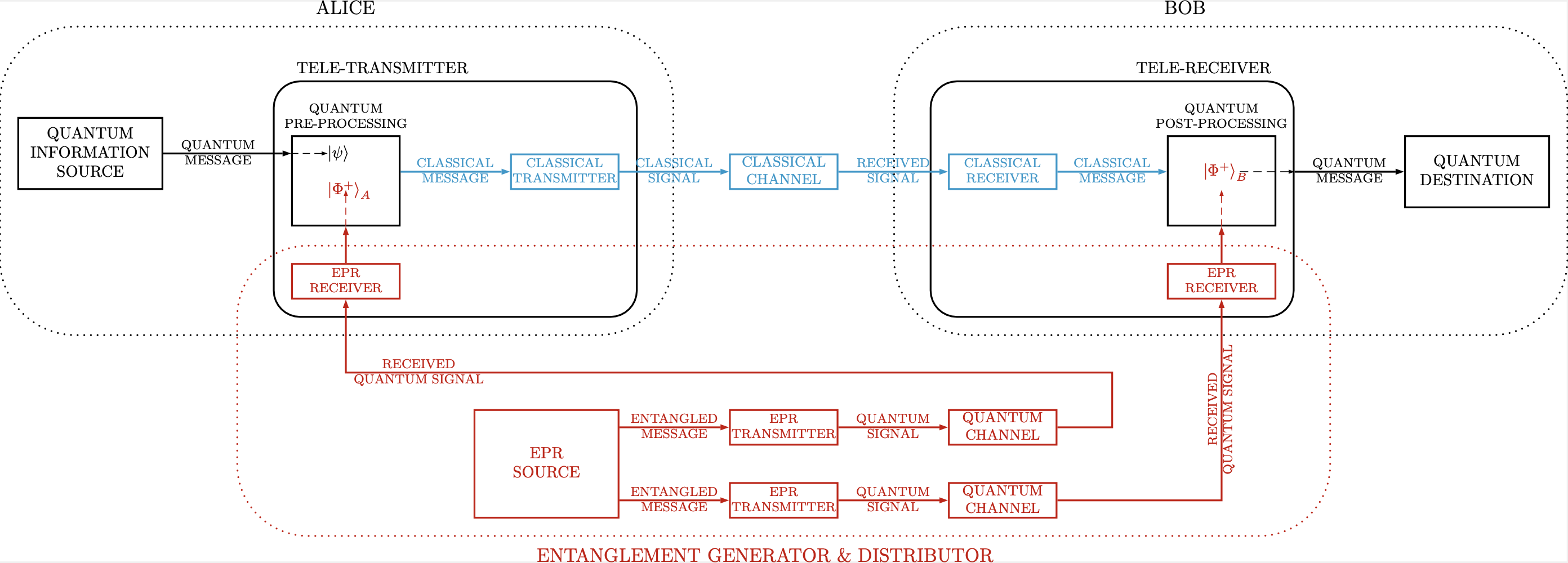}}
    \caption{Communication system model given in \cite{CacCalVan-20}. It can be straightforward applied to archetypes such as e-DQT exploiting direct modulation/demodulation.}
    \hrulefill
    \label{fig:x10}
\end{figure*}
Clearly, starting from the described source-destination link archetypes, it is possible to consider more complex network architectures, characterized by higher number of direct conversions or swapping nodes, accordingly to the specific network application and distance among the nodes. 

\begin{figure*}
    \hfill
    \begin{subfigure}{\linewidth}
        \includegraphics[width=\linewidth]{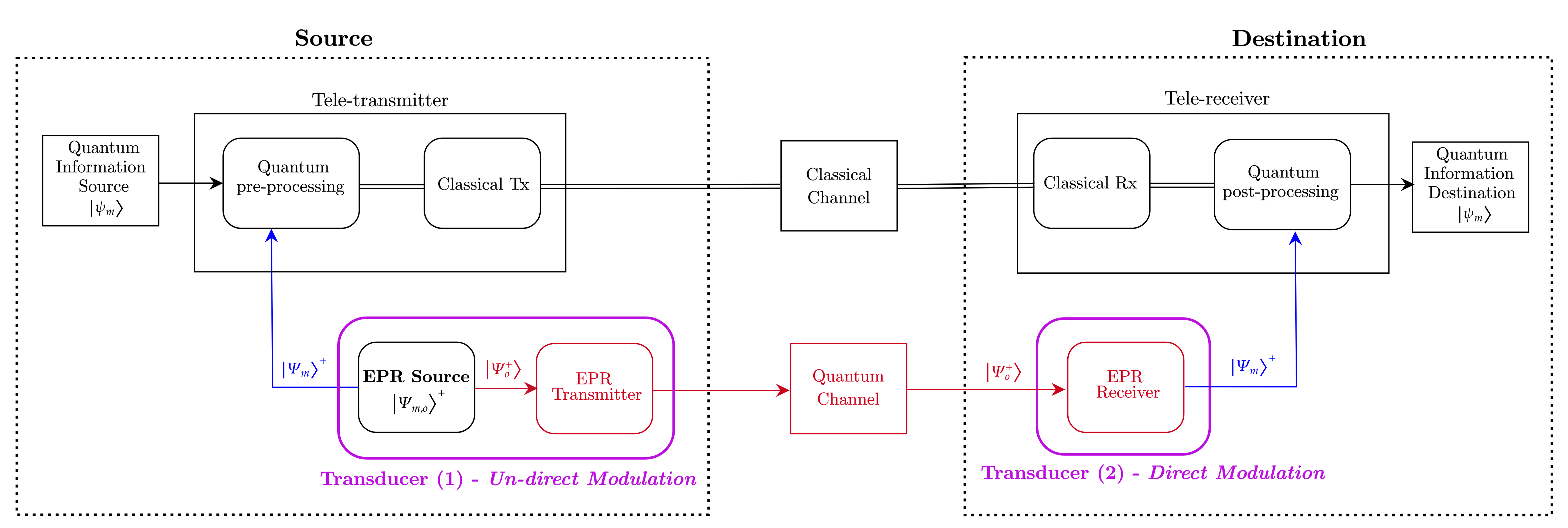}
        \caption{Communication system model for \textit{EGT Coupled with DQT} archetype.}
        \label{fig:x11.1}
    \end{subfigure}
    \hfill
    \begin{subfigure}{\linewidth}
        \includegraphics[width=\linewidth]{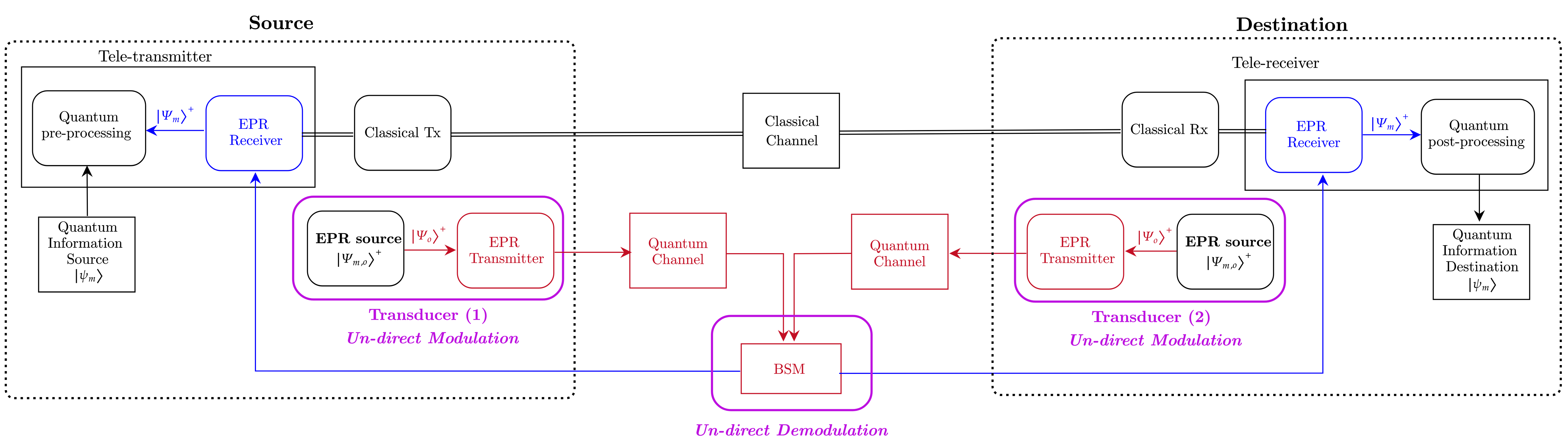}
        \caption{Communication system model for \textit{EGT Coupled with Swapping} archetype.}
        \label{fig:x11.2}
    \end{subfigure}
    \caption{Communication system models for archetypes exploiting un-direct modulation and/or un-direct demodulation.}
    \label{fig:x11}
    \hrulefill
    \end{figure*}

\section{Communication System Model}
\label{sec:06}

Stemming from the analysis developed in the previous sections, we are now ready to map quantum transduction into a functional block within a quantum communication system model.

More into details, the transducer can play a role reminiscent of the role played by a modulator at source side (or, equivalently, by a de-modulator at destination side) in a classical communication system model. Specifically, similarly to the classical domain where the modulation is the process of embedding an information signal into a second signal suitable for channel transmission \cite{OppWillYou-82}, in the quantum domain quantum transduction maps the qubit/ebit state into a quantum carrier suitable to be transmitted from the source through the channel.

However, while in the classical world there exists only one approach for implementing modulation/demodulation -- namely, the classical equivalent of direct modulation/demodulation described in Sec.~\ref{sec:01.1} -- in a quantum network direct modulation/demodulation is only one possibility. Indeed, in the quantum realm, we can distinguish between:
\begin{itemize}
    \item[-] \textit{direct modulation/demodulation} performed by a quantum transducer,
   \item[-] \textit{un-direct modulation} performed by a quantum transducer,
    \item[-] \textit{un-direct modulation} performed by the BSM.
\end{itemize}

Direct modulation/demodulation refers to the process of adapting the source output to the transmission channel (and vice-versa) at destination. This constitutes the functionality actually implemented by DQT on either the informational qubit or the ebit, as discussed in Secs.~\ref{sec:03.1} and \ref{sec:03.2} and represented in Fig.~\ref{fig:03}. In this sense, there exists a one-to-one mapping between a transducer and a modulator (demodulator) functional block within the classical Shannon communication system model \cite{Shan-48}. With this in mind and by elaborating further, we re-call the communication system model proposed in \cite{CacCalVan-20} and shown in Fig.~\ref{fig:x10}\footnote{In Fig.~\ref{fig:x10} ``Alice" and ``Bob" refers to source and destination nodes, respectively, and we refer the reader to \cite{CacCalVan-20} for an in-deep discussion of each block of the quantum communication model showed in the figure.}. The proposed model is a revision of the classical communication system model originally proposed by Shannon, which is refined to account for the specifics of quantum teleportation. In particular, in this model, the modulator/demodulator functional block must be mapped to (or included within) the \textit{EPR Transmitter} and \textit{EPR Receiver} blocks, and implemented by the quantum transducer. Furthermore, we can note that the model in Fig.~\ref{fig:x10} corresponds with a one-to-one mapping to the e-DQT archetype with entanglement generation at mid-point shown in Table~\ref{tab:04}. Furthermore, by inspection, it is possible to infer that it models as well any e-DQT archetype -- such as the e-DQT archetype with entanglement generation ``at source'' shown in Fig.~\ref{fig:x07.1} -- regardless of the location of entanglement generation, with minor adaptations.
Un-direct modulation refers to the functionality implemented by EGT, where the quantum transducer generates hybrid entanglement between microwave and optical modes. Notably, the optically-generated ebit is ready to be transmitted through the quantum channel, while the microwave ebit remains at the source for being successively exploited for quantum teleportation. In the above sense, there is no need to adapt the quantum carrier to the quantum channel, since the ebit generated in the optical domain is ``natively modulated'' to the quantum channel characteristics. Accordingly, we refer to this functionality implemented by EGT as \textit{un-direct modulation} -- being the modulation process ``virtually" performed via hybrid entanglement generation -- and we define EGT as a \textit{second generation} of quantum transducers. 
Stemming from the above considerations, we particularize in Fig.~\ref{fig:x11.1} the communication system model reported in Fig.~\ref{fig:x10} to the \textit{EGT Coupled with DQT} archetype showed in Fig.~\ref{fig:x07.2}. Accordingly, the transducer at the source performs un-direct modulation via EGT, while the transducer ad the destination keeps performing direct demodulation of the e-bits via DQT.

Finally, un-direct demodulation refers to the functionality implemented by the BSM within the \textit{EGT Coupled with Swapping}. For this archetype showed in Fig.~\ref{fig:x07.3}, the corresponding communication system model is schematically depicted in Fig.~\ref{fig:x11.2}. In this case, both transducers at source and destination implement EGT performing un-direct modulation as for the \textit{EGT Coupled with DQT}. But in this scenario also the demodulation is un-direct, since the demodulator block is not implemented by a transducer but rather by the BSM node, which distributes entanglement between non-interacting nodes by generating optical path entanglement. This implies that the process of ``receiving optical ebits and down-converting to microwave domain'' is fulfilled without the physical reception of the particle at neither the source or the destination. Therefore the EPR receiver block and the demodulation within is ``virtual'' \cite{CacCalVan-20}.

From the above, it becomes evident that all the possible source-destination link archetypes considered in Table~\ref{tab:04} can be analyzed and mapped into the  communication model proposed in \cite{CacCalVan-20}, although a one-to-one correspondence between modulation/demodulation and a quantum transducer (as happens in the classical domain) is not always possible due to the arising of indirect modulation/demodulation. In Table~\ref{tab:05}, we summarize the distinction between direct and un-direct modulation/demodulation.

\begin{table*}[t]
    
    \begin{tabularx}{\textwidth}{|
        >{\RaggedRight\arraybackslash}p{0.08\textwidth} |
        >{\RaggedRight\arraybackslash}p{0.08\textwidth} |
        >{\RaggedRight\arraybackslash}X |
        >{\RaggedRight\arraybackslash}p{0.18\textwidth} |
        }
        \toprule\toprule
        \multicolumn{2}{|c|}{\textbf{Quantum equivalent of modulation/demodulation}} 
        & \textbf{Functionality} 
        & \textbf{Transducer implementation} \\
        \hline
        \multirow{2}{*}{\textit{Direct}} 
         & Modulation 
         & mapping the qubit/ebit state available at the source into a quantum carrier suitable to be transmitted through the quantum channel 
         & \multirow{2}{*}{DQT} \\
        \cline{2-3}
         & Demodulation 
         & modulation-inverse operation
         &  \\
        \hline
        
        \multirow{2}{*}{\textit{Un-direct}} 
         & Modulation 
         & generation of an entangled resource where one ebit is natively mapped into a quantum carrier suitable to be transmitted through the quantum channel -- thus the modulation is \textit{virtual} -- whereas the other ebit is mapped into a carrier suitable for source processing/storing
         & EGT \\
        \cline{2-4}
         & Demodulation 
         & there is no direct demodulation of a photonic ebit carrier into a microwave one, but rather an indirect modulation where a Bell-State Measurement on the two photonic ebit carriers result into the two microwave ebit carries being entangled
         & there is no one-to-one mapping with a transducer\\
        \bottomrule\bottomrule
    \end{tabularx}%
    \caption{Direct vs un-direct modulation/demodulation.}
    \label{tab:05}
\end{table*}

\section{Discussion}
\label{sec:07}

The physical design and implementation of quantum transducers still faces several hardware challenges. And, beyond these inherent difficulties, further issues arise when considering their integration into large-scale communication systems.

Despite these open problems, quantum transduction remains one of the key functionalities for the implementation and deployment of quantum communication networks.

In this final section, we first highlight the transducer's fundamental role (even) in systems that operate natively at optical frequencies, where quantum transduction might be naively assumed as avoidable. Then, we discuss the challenges expected to emerge in future implementations, highlighting both the technological open issues as well as the architectural implications for scalable quantum networks.

\color{black}

\subsection{Intra-band Transduction}
\label{sec:07.1}

In this work, we analysed microwave-optical transduction with the aim of interconnecting distant superconducting quantum nodes via optical quantum channels.
However, as mentioned in Sec.~\ref{sec:01}, there exist several qubit platforms. Among them, trapped ions\cite{BliMoeDua-04, BlatWin-08}, quantum dots \cite{GaoFalEmr-12, HubReiAlb-18, SchReiBass-21} and spin-qubits \cite{TogChuTri-10} directly interact with optical photons. These qubit platforms emit entangled photons at visible/NIR wavelengths. Therefore, for long-distance entanglement distribution, it is necessary a frequency conversion to telecom frequencies. Indeed, O-band and C-band are commonly used for long-distance entanglement distribution \cite{AktFedKai-16, JorYehChe-24} with the ultimate goal of entanglement distribution on lit-fiber networks. \cite{BocEicPasc-98, RadDudZha-10, RanMaSla-10}. This frequency conversion from optical to optical frequencies is referred to as \textit{intra-band transduction} \cite{HanFuZou-21}. Therefore, also if some qubit platforms exploited for computation can spontaneously interact with optical photons, quantum transduction is inevitable for long-distance communications. And moreover, due to the intrinsically weak interactions between photons, a frequency converter for O- or C-band requires high power pump and coupling \cite{HanFuZou-21}.

In a nutshell, nowadays and in the near future, quantum transduction remains mandatory for quantum networks regardless of the considered underlying hardware platform. We choose to focus on the transduction between microwave and optical domains, since at the time of the manuscript writing superconducting technology constitutes the most promising platform for quantum computing. Indeed, superconducting quantum gates are fast \cite{ZhuJaaHe-21} with high-fidelity levels \cite{XuChuYua-20}, and their scalability allows to build quantum processors with hundreds of qubits \cite{PetrJahSae-24}.

\subsection{Challenges and Future Directions}
\label{sec:07.2}

At the hardware level, many challenges still need to be addressed for the realization of a quantum transducer device.

Firstly, as deeply discussed through this work, obtaining high conversion efficiency is one of the major hardware challenge. More into details, high efficiency requires a sufficiently large cooperativity, which critically relies on pump enhanced (linearized) coupling $g=g_0\sqrt{n_p}$, with $g_0$ and $n_p$ defined after Eq.~\eqref{eq:cooperartivity}.
Transducer devices can exhibit very low single-photon coupling rates ($g_0$), but the number of pump photons ($n_p$) injected into the device also plays a crucial role for an efficient conversion. For instance, integrated optomechanical devices can achieve very large coupling rates ($g_0 / 2\pi \sim 1$ MHz), but they typically operate with relatively few pump photons. In contrast, bulk electro-optical devices have much lower coupling rates (on the order of tens of Hz) but can accommodate a large number of injected photons. Integrated electro-optical devices typically have $g_0 / 2\pi \lesssim 1$ kHz while integrated optomechanical devices generally exhibit $g_0 / 2\pi \sim 10$ kHz, as summarized in in \cite{HanFuZou-21}. In general, the single photon nonlinearity $g_0$ depends on both the device design and the material coefficients. Careful device geometry engineering has already been done in many existing transducer designs to maximize the photon interaction allowing a more efficient conversion \cite{ChiBanMee-23, SamUtkSru-24}. Future improvement could be envisioned in novel material development \cite{XuLiWan-24} that can provide significantly larger nonlinear coefficients than those are currently being used, i.e. in LN, AlN, etc. \cite{GuoZouHoj-16, WanAliMarc-18,LuSurLiu-19}.

Beside the conversion efficiency, the added noise constitutes another critical challenge for the realization of a quantum transducer device. Within the frequency conversion, the noise can come from both thermal excitations in the microwave regime (which can be suppressed by high frequency operation $>$ GHz \cite{SahHeaRue-22, ZhaCheWill-25}) and the pump induced noise (additional heating, leakage photons and scattering, etc.). Therefore, the noise performance is coupled with pump photon number, and hence with the efficiency. Practically, it will be very important to develop new schemes to minimize the loss and heating at the device interfaces, improve the cooling power (e.g., via superfluid cooling \cite{FanZouCha-18, LiWuPac-25}), and at the same time reduce the optical photon scattering to the superconductors.

Compared with DQT, the EGT is advantageous in that it requires a lower pump photon number to maintain probability of entangled single-photon pair generation (a few percent) \cite{ZhoWanZhi-20}. Nevertheless, since the PDC process in non-deterministic, heralding is needed for EGT protocols which results in a reduced transduction rate.

For the considerations above, at the time of writing this manuscript the physics and hardware community is still primarily focused on improving quantum transduction performances according to the key parameters discussed so far. Consequently, the main technological challenges within these two research communities remain concentrated on the transduction platform itself.

\vspace{3pt}

Conversely, as for the communication engineering community, the main technological challenges switches to the integration of the quantum transducer within network archetypes and to more broader KPIs.

The implementation of the transducer must be scalable to address the transfer of multiple qubits between remote processors in parallel or generate and distribute multipartite entanglement \cite{DavCacCal-24, DavCacCal-2024}.
In particular, in the case of \textit{EGT Coupled with Swapping}, it will also be necessary to design precise synchronization control mechanisms. Indeed, such a link archetype requires the two optical photons reaching the BSM node being indistinguishable in ordered to successfully perform the swapping. This implies that hybrid EPRs generation should be temporally fine-tuned, by accounting and compensating for the delays introduced by the optical channels. Indeed, any timing discrepancies in the order of few dozens of picoseconds could compromise the communication performances of the archetype.

Furthermore, adopting a network-oriented perspective on quantum transduction calls for the definition of suitable performance metrics that go beyond the hardware-focused KPIs discussed above. Future investigations should explicitly address trade-offs such as fidelity versus entanglement distribution rate across remote nodes. Novel metrics -- such as end-to-end entanglement-distribution latency or overall scalability when complex topologies such as those arising with chains of quantum repeaters -- should be investigated.

The practical implementation of quantum transduction will constitute a cornerstone for large-scale quantum communications, while preserving compatibility with state-of-the-art computational technologies. This progress will pave the way toward distributed quantum computing (DQC) \cite{CalAmoFer-22} and Quantum Data Centers \cite{CacPelIll-25}, thereby unlocking entirely new classes of quantum applications. Novel and unprecedented functionalities are expected to emerge from a redefinition of the communication system model, driven by the introduction of new functional primitives -- such as quantum transduction via indirect modulation/demodulation -- no classical counterpart.

Indeed, the absence of a direct mapping onto the well-established foundations of the Classical Internet represents both a fundamental challenge in realizing the Quantum Internet and, simultaneously, one of its most intriguing opportunities.

With this work, we sought to highlight these aspects from a communication-engineering standpoint and to establish a foundation for the integration of quantum transducers into future quantum networks.

\color{black}

\bibliographystyle{IEEEtran}
\bibliography{bibliography.bib}
\end{document}